\documentclass[twocol]{ametsoc}
\pdfoutput=1 % To request/enforce running pdflatex for compilation

\usepackage{textcomp} % For \textdegree
\usepackage{verbatim} % For \begin{comment}
\usepackage{wasysym}  % For \apprge

\usepackage{gensymb} % for \degree in math mode

\usepackage{flushend} % To balance the last page (equal height on two cols)

\journal{jpo}

\bibpunct{(}{)}{;}{a}{}{,}

\title{On Oceanic Rogue Waves}

\authors{Francesco Fedele\correspondingauthor{Georgia Institute of Technology
Atlanta, GA 30332, USA.}}

\affiliation{School of Civil and Environmental Engineering, School of Electrical and Computer Engineering, Georgia Institute of Technology, Atlanta, GA, USA.}

\email{fedele@gatech.edu}

\abstract{
We propose a new conceptual framework for the prediction of rogue waves and third-order space-time extremes of wind seas that relies on the Tayfun (1980) and Janssen (2009) models coupled with Adler-Taylor (2009) theory on the Euler characteristics of random fields. As a specific application of this framework, extreme statistics of the 2007 Andrea rogue wave event are examined and verified with estimates from European Reanalysis (ERA)-interim data. In particular, the effects of nonlinear wave-wave interactions and space-time variability of the wave field are considered. A refinement of Janssen's (2003) theory suggests that in realistic oceanic seas characterized by short-crested multidirectional waves, homogeneous and Gaussian initial conditions become irrelevant as the wave field adjusts to a non-Gaussian state dominated by bound nonlinearities over time scales $t\gg t_{c}\approx0.13T_{0}/\nu\sigma_{\theta}$, where $T_{0}$, $\nu$ and $\sigma_{\theta}$ denote mean wave period, spectral bandwidth and angular spreading of dominant waves. For the Andrea storm, ERA-interim predictions yield $t_{c}/T_{0}\sim O(1)$ indicating that quasi-resonant interactions are negligible.  Further, the mean maximum sea surface height expected over the Ekofisk platform's area is higher than that expected at a fixed point within the same area. However, both of these statistics underestimate the actual crest height $h_{obs}\sim1.63H_s$ observed at a point near the Ekofisk site, where $H_s$ is the significant wave height. To explain the nature of such extreme, we account for both skewness and kurtosis effects and consider the threshold $h_{q}$ exceeded with probability $q$ by the maximum surface height of a sea state over an area in time.  We find that $h_{obs}$ nearly coincides with the threshold $h_{1/1000}\sim1.62H_s$ estimated at a point for a typical $3$-hour sea state, suggesting that the Andrea rogue wave is likely to be a rare occurrence in quasi-Gaussian seas.} 
\begin{document}

%% Necessary!
\maketitle

%%%%%%%%%%%%%%%%%%%%%%%%%%%%%%%%%%%%%%%%%%%%%%%%%%%%%%%%%%%%%%%%%%%%%
% MAIN BODY OF PAPER
%%%%%%%%%%%%%%%%%%%%%%%%%%%%%%%%%%%%%%%%%%%%%%%%%%%%%%%%%%%%%%%%%%%%%
%

%% In all cases, if there is only one entry of this type within
%% the higher level heading, use the star form: 
%%
% \section{Section title}
% \subsection*{subsection}
% text...
% \section{Section title}

%vs

% \section{Section title}
% \subsection{subsection one}
% text...
% \subsection{subsection two}
% \section{Section title}

%%%
% \section{First primary heading}

% \subsection{First secondary heading}

% \subsubsection{First tertiary heading}

% \paragraph{First quaternary heading}

\section{Introduction}

Rogue waves are unusually large-amplitude surface waves that appear from nowwhere in the open ocean. Evidences that such extremes can occur in nature are provided by the Draupner and Andrea events. In particular, the Andrea rogue wave was measured just past 00 UTC on November $9$ $2007$ by a LASAR system mounted on the Ekofisk platform in the North Sea in a water depth of $d=74$ m \citep{Magnusson2013}.  The Andrea wave has similar features of the Draupner freak wave measured by Statoil at a nearby platform ($d=70$ m) in  January 1995 \citep{haver2001evidences}.
Denoting the standard deviation of surface elevations by $\sigma$, the Andrea wave occurred during a sea state with significant wave height $H_{s}=4\sigma=9.2$ m, mean period $T_{0}=13.2$ s and wavelength $L_{0}=220$ m. 
The crest height is $h=15$ m ($h/H_{s}=1.63$) and the crest-to-trough height $H=21.1$ m ($H/H_{s}=2.3$) \citep{Magnusson2013}. The Draupner wave occurred during a 5-hour sea state with significant wave height $H_{s}=4\sigma=11.9$ m, mean period $T_{0}=13.1$ s and wavelength $L_{0}=250$ m. The crest height is $h=18.5$ m ($h/H_{s}=1.55$) and the crest-to-trough height $H=25.6$ m ($H/H_{s}=2.15$) \citep{haver2004possible,Magnusson2013}. In the last decade, the properties of the Draupner and Andrea waves have been extensively studied \citep{DystheKrogstad2008,Osborne2010,Magnusson2013,Bitner_Andrea2014,Dias2015} and references therein). Several physical mechanisms have been proposed to explain the occurrence of such giant waves \citep{Kharif2003_a}, including the two competing hypotheses of nonlinear focusing due to third-order quasi-resonant wave-wave interactions (\cite{Janssen2003}), and purely dispersive focusing of second-order waves \citep{Fedele2009,Fedele2008a}. 

\begin{figure*}[t]
\centering\includegraphics[width=0.75\textwidth]{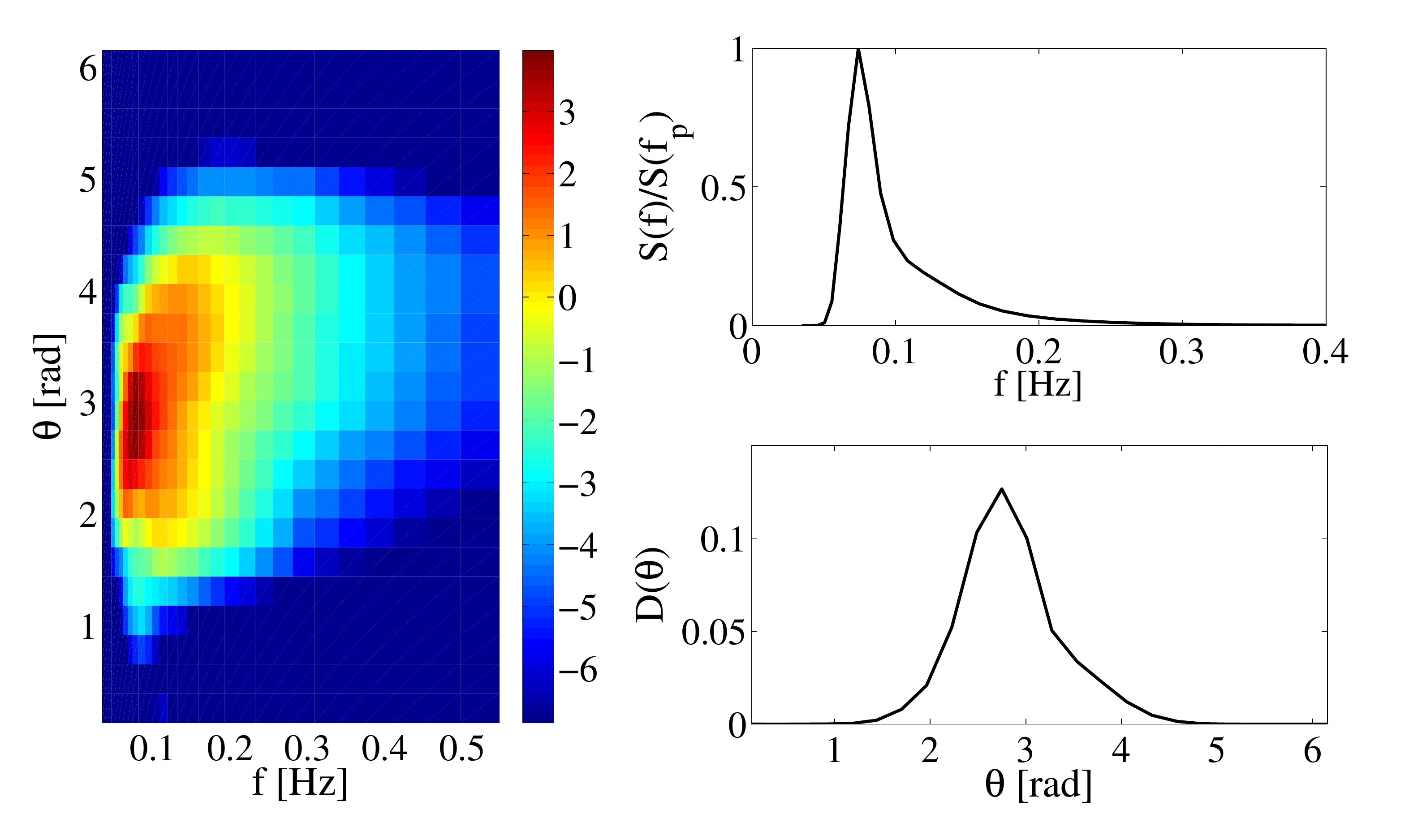}
\caption{Draupner storm: ERA-interim (left) directional spectrum (log scale)
at the Draupner site ($58.2\degree$ N, $2.5\degree$ E) 
at the time of maximum development of the storm (Jan 2nd 1995 00:00 UTC) and (top-right) wave frequency spectrum $S(f)/S(f_{p})$ and (bottom-right)
angular dispersion $\sigma^{2}D(\theta)=\int S(f,\theta)df$, where
$\sigma$ is the standard deviation of surface elevations and $f_{p}$
the dominant frequency. Direction $\theta=0$ means going to the north
and $\theta=\pi/2$ to the east (Oceanographic convention).}
\label{FIG1} 
\end{figure*}

For instance, recent studies propose the hypothesis that the Draupner wave
occurred in crossing seas (see e.g. \cite{onorato2010freak}). These suggest that angles lying in the range  $\sim10\text{\textdegree}-30{^\circ}$ between two dominant sea directions are likely to lead to rogue-wave occurrences induced by quasi-resonant wave-wave interactions.
However, \cite{adcock2011did} reported that the hindcast from the
European Centre for Medium-Range Weather Forecasts shows swell waves
propagating at approximately 80\textdegree{} to the wind sea. \cite{adcock2011did} also argued that the Draupner wave occurred due to the crossing of two almost orthogonal wave groups in accord with
second-order theory. This would explain the set-up observed under the large wave \citep{Walker2004} instead of the second-order set-down normally expected \citep{LonguetHiggins1964}. However, there is no evidence of significant swell components nearby the platform as clearly seen from Fig. 2 in \cite{adcock2011did} and Fig. \ref{FIG1} here, which shows the ERA-interim wave directional spectrum at the Draupner site. Further, in accord with Boccotti's (2000) quasi-determinism (QD) theory the probability that two different wave groups cross at the same point at the apex of their development is much smaller than the probability that one of the two groups focuses at the same point\nocite{Boccotti2000}. One can also argue that reflection and diffraction from the platform may cause the observed set-up.  However, the Draupner measurements were made from a bridge connecting two space frame structures. The structural members are relatively small, likely a meter in diameter.  The preceding greatly lessens the chances for platform interference from spray, reflections or diffractions.
%Clearly, one can also argue that the observed set-up is an indication that measurements may be corrupted. 
Clearly, the Draupner wave appears fundamentally different from a typical expected extreme wave because of the observed set-up of the mean sea level (MSL) below the large crest. However, the estimation of the MSL from measurements is ill-defined and thus may be not robust. Indeed, we note that \cite{Walker2004} estimated the mean sea level by low-pass filtering the measured time series of the wave surface with frequency cutoff $f_c\sim f_p/2$, where $f_p$ is the dominant frequency. Clearly, the mean surface and wave fluctuations are nonlinearly coupled and feed energetically into each other. As a result, the low-pass filtered mean surface is a mix of the two components. If the time series is not long enough for a statistically significant estimation of wave-wave interactions, the observed set-up could be the manifestation of the large crest segment that extends above the adjacent lower crests. In this work, we will not dwell on the Draupner wave. This calls for further studies and numerical simulations of the sea state in which it occurred to clarify the physics and robustness of the observed set-up below the large crest. These issues are secondary and are beyond the scope of this paper. Indeed, we will capitalize on recent numerical simulations of the Andrea rogue wave \citep{Bitner_Andrea2014,Dias2015} in order to study the statistical properties and space-time extremes of the sea state in which it occurred. 

Other studies propose third-order quasi-resonant interactions and associated modulational instabilities as mechanisms for rogue wave formation (e.g. \cite{Janssen2003, Osborne2010}). Such nonlinear effects cause the statistics of weakly nonlinear gravity waves
to significantly differ from the Gaussian structure of linear seas, especially in long-crested, or unidirectional seas (\cite{Janssen2003,Fedele2008a,Onorato2009,ShemerGJR2009,Toffoli2010,fedeleNLS}). 
%The wave field near a large crest is that of a breather (\cite{Peregrine1983,Osborne2000},\cite{Ankiewicz2009}).
The relative importance of such nonlinearities and the increased occurrence of large waves can be measured by the excess kurtosis 
\begin{equation}
\lambda_{40}=\frac{\left\langle \eta^{4}\right\rangle }{\left\langle \eta^{2}\right\rangle ^{2}}-3,
\end{equation}
of the mean-zero surface elevation $\eta$, where brackets denote statistical average. This integral statistic is defined by \cite{Janssen2003} as $C_{4}=\lambda_{40}/3$. In general,
$C_{4}=C_{4}^{d}+C_{4}^{b}$ and it comprises a dynamic component $C_{4}^{d}$ due to nonlinear wave-wave interactions (\cite{Janssen2003}) and a bound contribution $C_{4}^{b}$ induced
by the characteristic crest-trough asymmetry of ocean waves (\cite{Tayfun1980,TayfunLo1990,TayfunFedele2007,Fedele2009,Fedele2008a, Janssen2009}).
For deep-water long-crested seas with a Gaussian-shaped spectrum and within the framework of the higher
order compact Zakharov (cDZ) equation (\cite{Dyachenko2011}), \cite{Fedele2014}
showed that, correct to $O(\nu^{2})$ in spectral bandwidth, the dynamic
excess kurtosis monotonically increases in time toward the asymptotic value
\[
C_{4,cDZ}^{d}=C_{4,NLS}^{d}\left(1-\frac{4\sqrt{3}+\pi}{8\pi}\nu^{2}\right)\approx C_{4,NLS}^{d}\left(1-0.40\nu^{2}\right),
\]
where 
\begin{equation}
C_{4,NLS}^{d}=BFI^{2}\frac{\pi}{3\sqrt{3}}\label{C4R-1}
\end{equation}
is the dynamic excess kurtosis of unidirectional narrowband
waves in deep water in accord with the one-dimensional (1-D) nonlinear Schr\"odinger (NLS)  equation (\cite{Janssen2006}). Note that Eq.~\eqref{C4R-1} is also valid for the Dysthe (1979) equation as the associated asymmetric spectral broadening is not captured by the assumed symmetric Gaussian spectrum. The Benjamin-Feir index $BFI={\sqrt{2}\mu}/{\nu}$, with $\mu$
denoting an integral measure of wave steepness defined later in Eq.~\eqref{mua} and the spectral bandwidth $\nu$ is given in Eq.~\eqref{nu}. 
Clearly, $C_{4,cDZ}^{d}$ is smaller than $C_{4,NLS}^{d}$, especially as the spectral bandwidth widens. 
This is consistent with the result that in accord with cDZ the linear growth rate of a subharmonic
perturbation reduces with respect to the NLS counterpart for waves with broader spectra. Indeed, for fixed wave steepness, the initial-stage growth of instabilities away from a Stokes wave is attenuated as the spectral bandwidth increases (\cite{Alber1978,Crawfordetal1981}).
The late-stage evolution of modulation instability leads to breathers that can cause large waves \citep{Peregrine1983,Osborne2000,Ankiewicz2009}, especially in unidirectional waves. Indeed, in this case energy is ''trapped'' as in a long wave-guide. For small wave steepness and negligible dissipation, quasi-resonant interactions are effective in reshaping the wave spectrum, inducing larger breathers via nonlinear focusing before breaking occurs (\cite{ShemerGJR2009,Onorato2009,Chabchoub2011,Chabchoubc2012}). \cite{ShemerPoF2013} pointed out that wave breaking is inevitable for $\mu>0.1$, and breathers can be observed experimentally only at sufficiently small values of wave steepness ($\sim0.01-0.09$)  (\cite{Chabchoub2011,Chabchoubc2012}, see also \cite{Shemer2014}). Further, they also noted that 'breather does not breath' and its amplification is smaller than that predicted by the NLS, in accord with the numerical studies of the Euler equations (\cite{SlunyaevShrira2013,Slunyaev2013PRE}). However, typical oceanic wind seas are not 1-D but short-crested multidirectional wave fields. Hence, nonlinear focusing due to modulational effects is diminished since energy can spread directionally (\cite{Onorato2009,Toffoli2010}). 
 
\begin{figure*}[t]
\centering\includegraphics[width=0.75\textwidth]{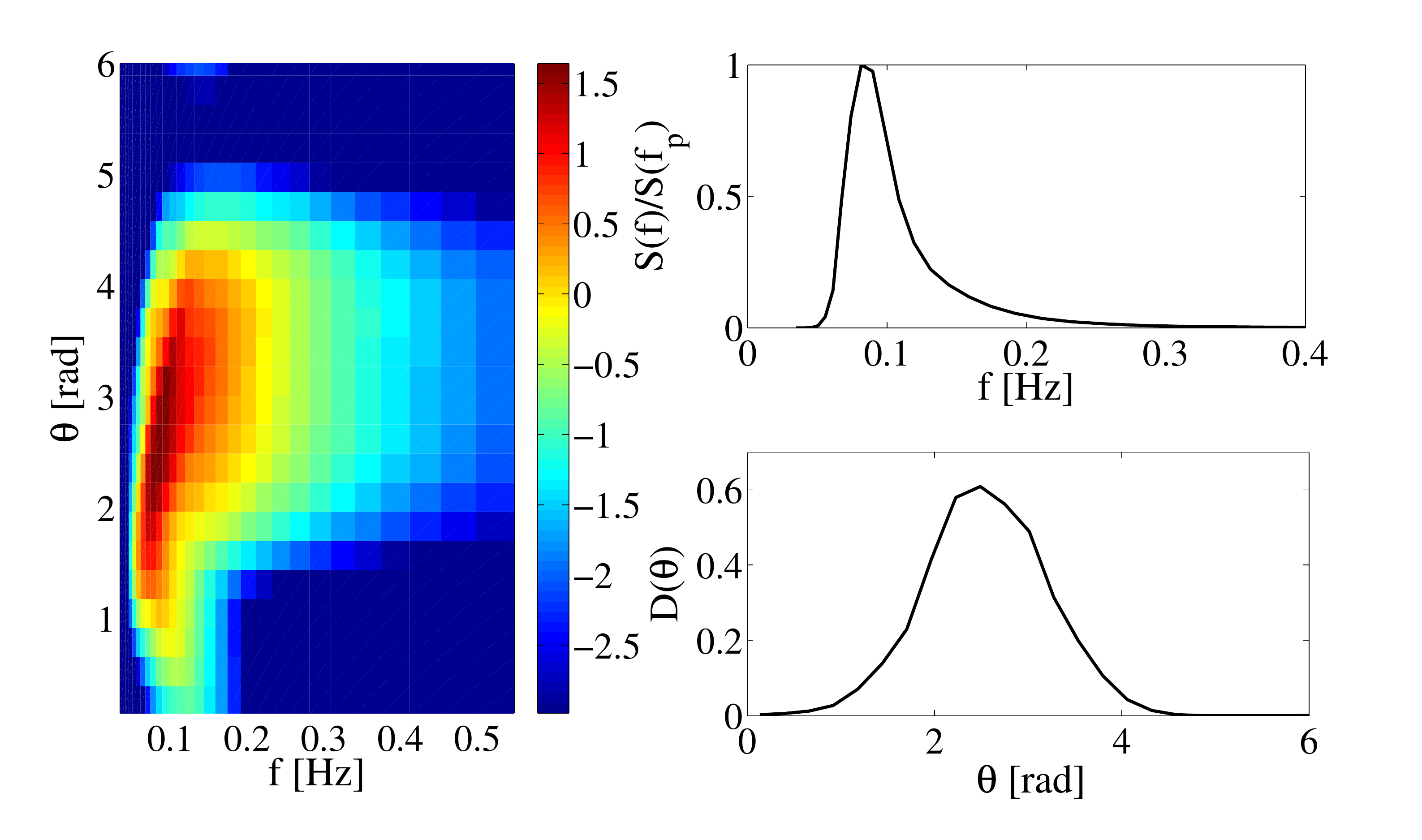}
\caption{Andrea storm: ERA-interim (left) directional spectrum (log scale)
at the Ekofisk site ($56.3\degree$ N, $3\degree$ E) 
at about the time of the Andrea wave event (Nov 9th 2007, 00:00 UTC) 
and (top-right) wave frequency spectrum $S(f)/S(f_{p})$ and (bottom-right)
angular dispersion $\sigma^{2}D(\theta)=\int S(f,\theta)df$, where
$\sigma$ is the standard deviation of surface elevations and $f_{p}$
the dominant frequency. Direction $\theta=0$ means going to the north
and $\theta=\pi/2$ to the east (Oceanographic convention).}
\label{FIG2} 
\end{figure*}

The sea state of the Andrea wave was short-crested (see Fig. \ref{FIG2}), and modulation instabilities may have played an insignificant role in the wave growth \citep{Alber1978,Crawfordetal1981}. Further, the actual water depth to wavelength ratio, namely $d/L_{0}\sim0.3$, suggests that waves were in transitional regime where modulation instabilities, if any, are further attenuated (see, e.g. \cite{Toffoli2009}).
Recently, \cite{Tayfun2008} arrived at similar conclusions based
on the analysis of data  from the North Sea. His results indicate that large 
time waves (measured at a given point) result from the constructive
interference (focusing) of elementary waves with random amplitudes
and phases enhanced by second-order non-resonant interactions. Further,
the surface statistics follow the Tayfun (1980) distribution\nocite{Tayfun1980} in agreement with observations (\cite{TayfunFedele2007,Fedele2008a,Tayfun2008,Fedele2009}). This is confirmed by a recent data quality control and statistical analysis by \cite{Christou2014}
of single-point measurements from fixed sensors mounted on offshore
platforms, the majority of which were recorded in the North Sea. The
analysis of an ensemble of 122 million individual waves revealed 3649
rogue events, concluding that rogue waves observed at a point in time are merely rare events induced by dispersive
focusing.

More recent studies on the statistics of extreme ocean waves provide both theoretical and experimental evidences that the expected maximum sea surface height over an area in time (space-time extreme) is larger than that expected at a fixed point (time extreme), especially in short-crested multidirectional seas (\cite{forristall2011,forristall2015,Fedele2012,fedele2013,Barbariol2014}). The occurrence of an extreme in Gaussian fields is analogous to that of a big wave that a surfer searches and eventually finds (\cite{Baxevani2006}). If the surfer searches a large area, his chances of encountering the largest crest obviously increase. Indeed, a large number of radar signatures of likely rogue waves have been identified from satellite data in the context of the MaxWave project  \citep{MAXPROJECT2008}. However, the relation between space-time extremes and rogue waves is still unclear.  

The preceding review provides the principal motivation for introducing the theory of stochastic space-time extremes and applying it to study the space-time properties of sea states in which rogue waves occur.  In this work, we only focus on the Andrea rogue event as the Draupner wave requires further studies on the observed set-up \citep{Walker2004}. We aim at presenting a new conceptual framework for the prediction of large waves based on the Tayfun (1980) and Janssen (2009) models coupled with Adler-Taylor (2009) theory on the Euler-characteristics of random fields. The framework so developed can be applied to study both second and third-order nonlinear effects and space-time properties of sea states where rogue waves are expected to occur. In particular, the new stochastic model describes the statistical structure of the wave surface using a Gram-Charlier type distribution for crest heights formulated in \citep{TayfunFedele2007}. This is able to capture the vertical crest-trough asymmetry induced by second-order nonlinearities measured by skewness \citep{Tayfun1980} as well as intermittency effects due to third-order nonlinearities measured by kurtosis \citep{JanssenJFM2009}. 
Moreover, the statistical properties of local waves in space and time are related to higher order moments of the wave directional spectrum capitalizing on Adler-Taylor (2009) theory. As a result, we are able to define the probability structure of the random variable $\eta_{\mathrm{max}}$ representing the maximum third-order nonlinear surface wave height over an area in time. In accord with the new model, we will show that statistical averages of the maximum $\eta_{\mathrm{max}}$ do not explain rare occurrences of large oceanic waves, which instead need be interpreted using quantile-type approaches. In particular, we will consider the threshold $h_{q}$ exceeded by $\eta_{\mathrm{max}}$ with probability $0\le q \le1$ and the conditional mean $\overline{h}_q=\left\langle\eta_{\mathrm{max}}|\eta_{\mathrm{max}}>h_{q}\right\rangle$, where brackets denote statistical average. The statistic $h_q$ is a generalization to space-time maxima of the threshold $c_{n}$ exceeded by the crest height of the $1/n$ fraction of largest waves observed at a point in time \citep{TayfunFedele2007}). Similarly, $\overline{h}_{q}$ generalizes the crest height average $c_{1/n}$ of the $1/n$ fraction of largest waves \citep{TayfunFedele2007}). 

The remainder of the paper is organized as follows. First, the essential elements of Janssen's (2003) formulation for the excess kurtosis of directional or short-crested seas are presented \citep{fedele2014kur}. This is followed by a review of the theory of Euler characteristics of random fields
(\cite{adler1981geometry}), space-time extremes \citep{Fedele2012}
and associated stochastic wave groups \citep{Fedele2009}. We then introduce a new stochastic model for the prediction of space-time extremes that accounts for both second and third-order nonlinearities. 
As a specific application here, capitalizing on the ERA-interim reanalysis (\cite{ERA}) and numerical simulations of the Andrea sea state \citep{Bitner_Andrea2014,Dias2015}, the extreme statistics of the Andrea rogue-wave event is examined in detail. 
In concluding, we discuss the implications of these results on rogue-wave predictions.

\section{Excess kurtosis of short-crested seas }

\cite{fedele2014kur} revisited Janssen's (2003) formulation for the
total excess kurtosis $C_{4}$ of weakly nonlinear gravity waves in
deep water. This comprises a dynamic component $C_{4}^{d}$ due to
nonlinear wave-wave interactions (\cite{Janssen2009}) and a bound contribution $C_{4}^{b}$ (\cite{JanssenJFM2009}).
For waves that are narrowband and characterized by
a Gaussian type directional spectrum, $C_{4}^{d}$ is expressed as
a six-fold integral that depends on time $t$, $BFI$ and the parameter 
\begin{equation}
R=\frac{\sigma_{\theta}^{2}}{2\nu^{2}},\label{RR}
\end{equation}
which is a dimensionless measure of short-crestedness of dominant waves, with $\nu$
and $\sigma_{\theta}$ denoting spectral bandwidth and angular spreading
(\cite{Janssen2009,Mori2011}). The associated excess kurtosis growth
rate can be solved analytically for narrowband waves (\cite{fedele2014kur}, see also Appendix~A). It is found that in the focusing regime ($0<R<1$) the dynamic
excess kurtosis initially grows attaining a maximum $C_{4,\max}$ at the
intrinsic time scale 
\begin{equation}
\tau_{c}=2\pi\nu^{2}\frac{t_{c}}{T_{0}}=\frac{1}{\sqrt{3R}},\qquad\mathrm{or}\qquad\frac{t_{c}}{T_{0}}\sim\frac{0.13}{\nu\sigma_{\theta}}\label{tau}
\end{equation}
given by the least-squares fit 

\begin{equation}
\frac{C_{4,\max}^{d}(R)}{BFI^{2}}\approx\frac{b}{(2\pi)^{2}}\frac{1-R}{R+bR_{0}},\qquad\qquad0\leq R\leq1,\label{fit}
\end{equation}
where $R_{0}=\frac{3\sqrt{3}}{\pi}$ and $b=2.48$. Eventually the
excess dynamic kurtosis tends monotonically to zero as energy spreads
directionally, as in the numerical simulations of \cite{ShriraGRL2009}.
In the defocusing regime ($R>1$), the dynamic excess kurtosis is always
negative. It attains a minimum at $t_{c}$ given by (\cite{Janssen2009})
\begin{equation}
C_{4,\mathrm{min}}^{d}\left(\frac{1}{R}\right)=-RC_{4,\max}^{d}(R),\qquad\qquad0\leq R\leq1.\label{minC}
\end{equation}
and then tends to zero in the long time. Thus, the present theoretical
predictions indicate a decaying trend for the dynamic excess kurtosis
over large times.

\begin{figure*}[t]
\centering\includegraphics[width=\textwidth]{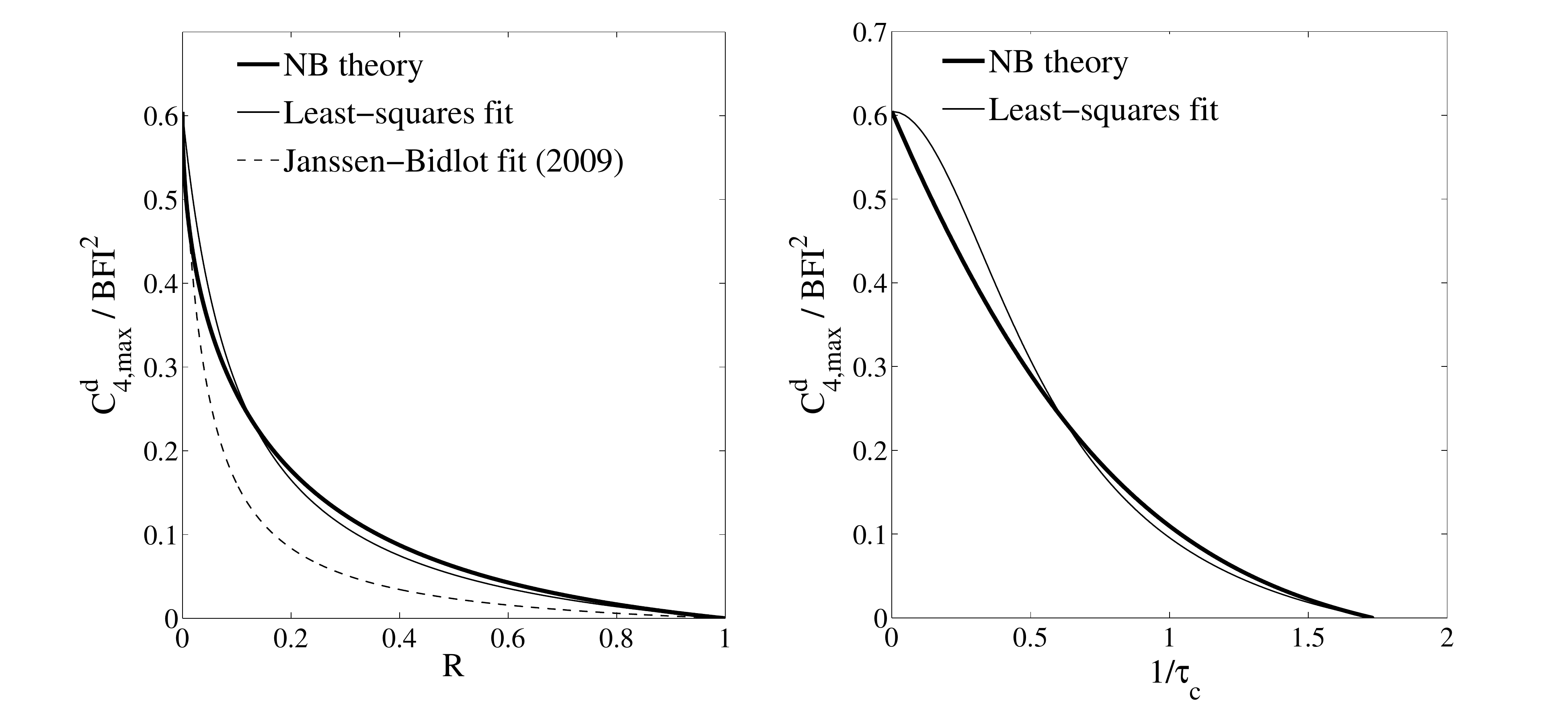}
\caption{Maximum dynamic excess kurtosis $C_{4,\mathrm{max}}^{d}$ as a function
of (left) $R$ and (right) $1/\tau_{c}$: (bold curve) present theoretical
prediction, (thin curve) least-squares fit from Eq. (\ref{fit}) ($b=2.48$)
and (dashed curve) Janssen-Bidlot (2009) fit ($b=1$).}
\label{FIG3} 
\end{figure*}

For time scales $t\apprge10t_{c}$, a cold start with initial homogeneous and Gaussian 
conditions become irrelevant as 
the wave field tends to a non-Gaussian state dominated by bound nonlinearities 
as the total kurtosis of surface elevations asymptotically approaches the value 
represented by the bound component (\cite{Shrira2013_JFM,Shrira2014_JPO}).
In typical oceanic storms where dominant waves are characterized with
$\nu\sim0.2-0.4$ and $\sigma_{\theta}\sim0.2-0.4$, this adjustment
is rapid since the time scale $t_{c}/T_{0}\sim O(1)$ with $T_{0}\sim10-14$
s and the dynamic kurtosis peak is negligible compared to the bound
counterpart. For time scales of the order of or less than $t_{c}$,
the dynamic component can dominate and the wave field may experience
rogue wave behavior induced by quasi-resonant interactions (\cite{Janssen2003}).
However, one can argue that the large excess kurtosis transient observed
during the initial stage of evolution is a result of the unrealistic
assumption that the initial wave field is homogeneous Gaussian 
whereas oceanic wave fields are usually statistically inhomogeneous both in space and time. In the left panel of Fig.~\ref{FIG3}, the preceding approximation in Eq.~\eqref{fit}
is compared against the theoretical $C_{4,\max}^{d}$ for narrowband
waves (\cite{fedele2014kur}, see also Appendix~A). Evidently, the
latter is slightly larger than the maximum excess kurtosis derived
by \cite{Janssen2009}, who have also used the fit in Eq.~\eqref{fit} but with $b=1$.
Their maximum follows by first taking the limit of the excess kurtosis
at large times and then solving the associated six-fold
integral (\cite{fedele2014kur}). Clearly, the dynamic excess kurtosis
should vanish at large times. Janssen (personal communication, 2014\nocite{Janssen2014})
confirmed that Eq.~\eqref{Cmax} holds and provided an alternative
proof that the large-time $C_{4}^{d}$ tends to zero using
complex analysis. 

Further, in the focusing regime ($R<1,\tau_{c}<1/\sqrt{3}$), from
Eq. (\ref{fit})
\begin{equation}
\frac{C_{4,\max}^{d}(\tau_{c})}{BFI^{2}}\approx\frac{b}{(2\pi)^{2}}\frac{-1+3\tau_{c}^{2}}{1+3bR_{0}\tau_{c}^{2}}.\label{fit2}
\end{equation}
Clearly, the maximum kurtosis becomes larger for longer time scales
$\tau_{c}$, as illustrated in the right panel of Fig.~\ref{FIG3}.
In the defocusing regime ($R>1$,$\tau_{c}>1/\sqrt{3}$), the dynamic
excess kurtosis is negative and its minimum value $C_{4,\mathrm{min}}^{d}$
can be computed from Eq.~\eqref{minC}.
This result holds for deep-water waves. Drawing on \cite{JanssenOnorato2007} and \cite{Janssen2009}, the extension to intermediate waters of depth $d$ follows by redefining the Benjamin-Feir Index as $BFI_{S}^2=\alpha_{S} BFI^2$, where the depth factor $\alpha_{S}$ depends upon the dimensionless depth $k_0d$, where $k_0$ is the dominant wavenumber (see Appendix~A). In this work we choose $k_0$ as the mean wavenumber $k_{m}$ corresponding to the mean spectral frequency $\omega_{m}=m_{001}/m_{000}$, where $m_{ijk}$ are spectral moments (see Appendix~B).  In the deep-water limit $\alpha_{S}$ becomes 1, and $BFI_{S}$ reduces to the usual definition of $BFI$ (\cite{Janssen2003}).
As the dimensionless depth $k_0 d$ decreases, $BFI_S^2$ reduces and it becomes negative for $k_0 d<1.363$ and so the dynamic excess kurtosis.
%Also the parameter a needs adjustment for the shallow water case, but this has not been introduced yet.

\begin{figure*}[t]
\centering\includegraphics[width=1.1\textwidth]{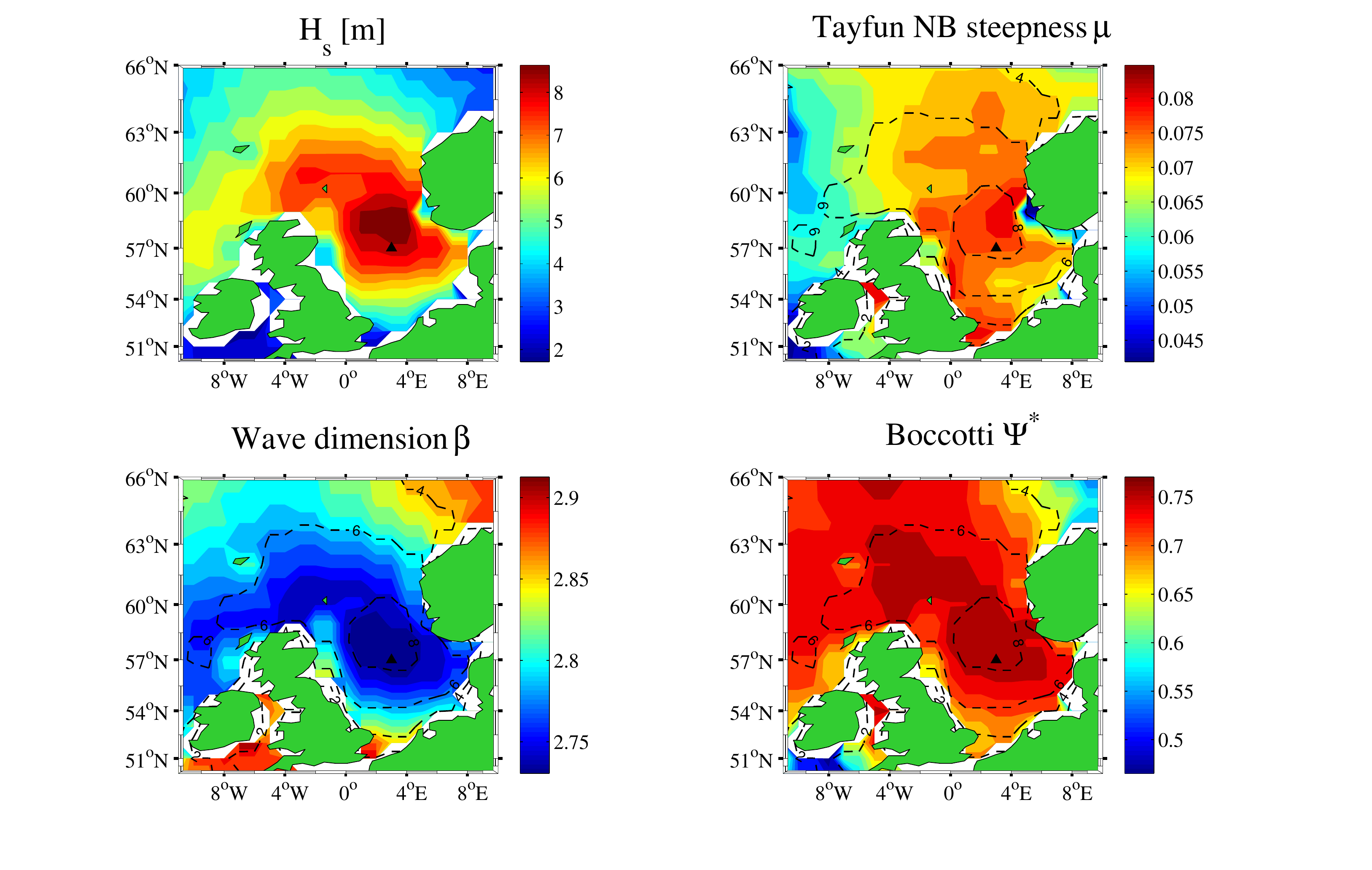} 
\caption{ERA-interim reanalysis at the peak of the Andrea storm. Top panels: (left) significant wave height
$H_{s}=4\sigma$ and (right) Tayfun NB wave steepness $\mu_S$ (Eq. (\ref{mua-1})).
Bottom panels: (left) wave dimension $\beta$ (Eq. (\ref{beta}))
and (right) narrowbandedness Boccotti parameter $\psi^{*}$. Dashed
lines are $H_{s}$ contours. The triangle symbol indicates the Ekofisk site position.}
\label{FIG4} 
\end{figure*}
Thus, we expect that third-order quasi-resonant interactions do not play any role in the formation of large waves in realistic oceanic seas. However, the effects of bound nonlinearities on skewness and kurtosis should be accounted for in the extreme value analysis.
In this regard, \cite{JanssenJFM2009} deeply investigated the properties of weakly nonlinear water waves within the Hamiltonian formalism developed by \cite{Krasitskii1994}. Further, he derived general expressions for skewness and kurtosis of the wave surface valid for finite depths and arbitrary spectra. In particular, the canonical transformation for a third-order narrowband (NB) wavetrain can be evaluated explicitly and \cite{JanssenJFM2009} gives a simple expression for the NB bound kurtosis given in Eq. (\ref{Cb}) of Appendix~A. 

\cite{Bitner_Andrea2014} and \cite{Dias2015} performed Monte Carlo simulations of the hindcasted sea state in which the Andrea wave occurred using the irrotational and inviscid Euler equations. The estimated kurtosis is mainly due to non-resonant wave-wave interactions and  $C_4=C_4^b\sim0.033$, or $\lambda_{40}=\lambda_{40}^b=3C_4\sim0.1$. This is in fair agreement with  Janssen's NB estimate $\lambda_{40}^{b,NB}=0.09$ from Eq. (\ref{Cb}) using ERA-interim data. \cite{Bitner_Andrea2014} also reported a larger kurtosis ($\sim0.35$) several hours before the Andrea wave occurrence. The actual laser measurements at Ekofisk give instead oscillating values within the range  $0-0.3$ indicating unstable estimates of kurtosis as they are based on short 20-min time series (\citep{Magnusson2013}). 

\begin{figure*}[t]
\centering\includegraphics[width=1.07\textwidth]{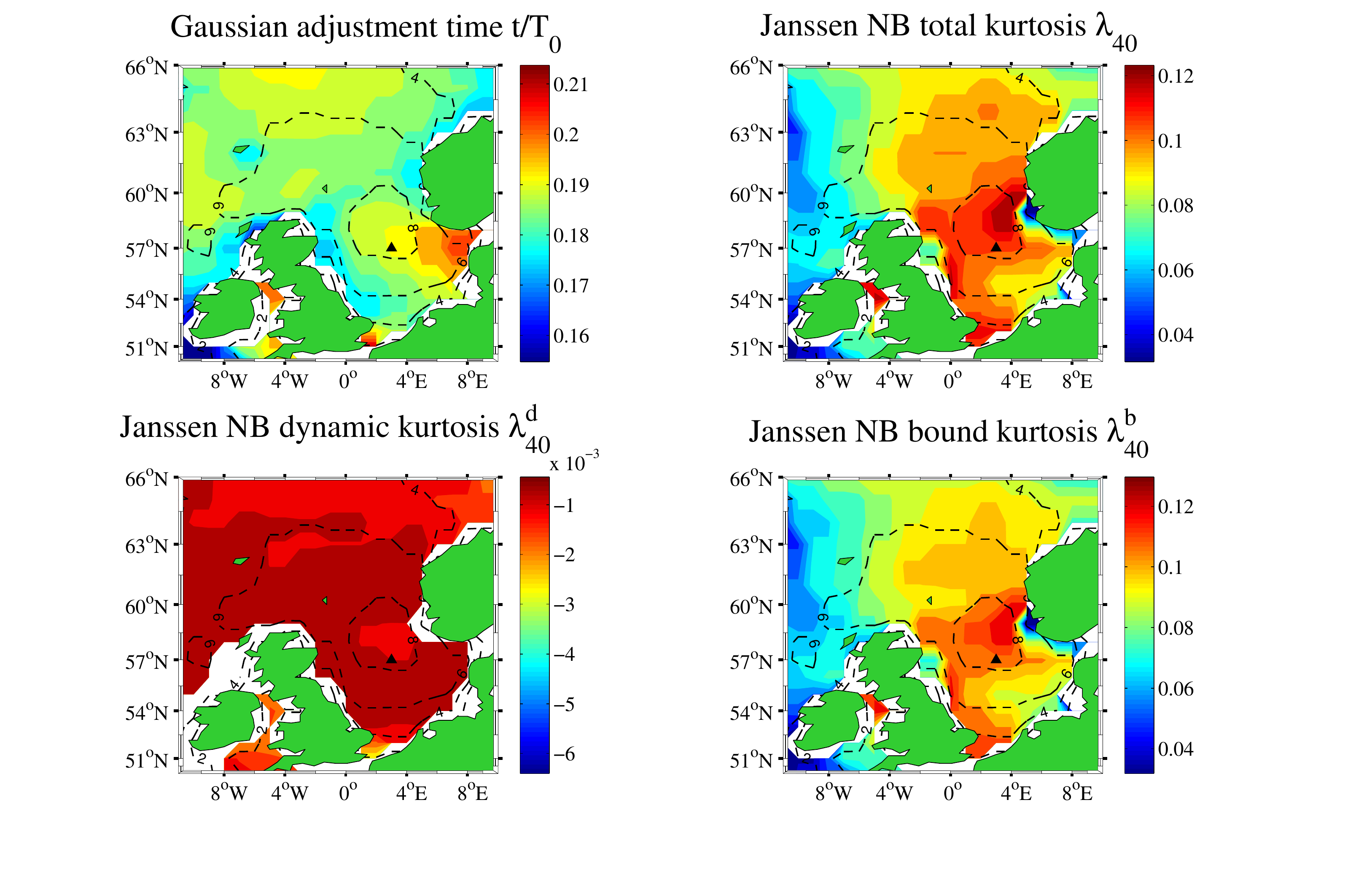} 
\caption{ERA-interim reanalysis at the peak of the Andrea storm. Top panels: (left) Gaussian adjustment
time $t_{c}/T_{0}$ (Eq. (\ref{tau})) and (right) total NB kurtosis
$\lambda_{40}=\lambda_{40}^{d}+\lambda_{40}^{b}$. Bottom panels: (left) NB dynamic kurtosis $\lambda_{40}^{d}$ (Eq. (\ref{fit})) and (right) NB bound kurtosis $\lambda_{40}^{b}$. Dashed lines are $H_{s}$ contours [m]. 
The triangle symbol indicates the Ekofisk site position.}
\label{FIG5} 
\end{figure*}

Drawing on the ERA-interim reanalysis data, we now consider the hindcasted sea state during which the Andrea wave occurred (UTC 00 on Nov 9th, 2007). 
The top panel on the left of Fig.~\ref{FIG4} shows the spatial distribution of significant wave
height. The maximum $H_{s}$ is about $8.3$ m, which is smaller than $9.2$ m actually observed (\cite{Magnusson2013}). It is well known that ERA-interim underestimates peak values and predicts
broader directional spectra because of the low space and time resolution
of the data. Indeed, wave parameters are solved every $6$ hours and the grid cell areal size is $A_0\sim100^{2}$ $\mathrm{km^{2}}$ with 60 vertical levels (\cite{ERA}). Nevertheless, such predictions
provide leading order estimates of  the sea-state parameters that can be refined further in future studies, 
using forecast models with higher resolution (\cite{Sonia}). The top panels of Fig.~\ref{FIG5} shows the Gaussian adjustment time $t_{c}/T_{0}$ and the Janssen NB total excess kurtosis $\lambda_{40}=3C_{4}$. The NB dynamic component $\lambda_{40}^d=3C_{4}^d$ from Eq.~\eqref{fit} and the bound counterpart $\lambda_{40}^b=3C_{4}^b$ from Eq.~\eqref{Cb} are shown in the bottom
panels of the same figure. At the Ekofisk location the water depth is $d=74$ m, so $k_0 d\sim3$ and $\alpha_S=0.58$. 
As a result the dynamic kurtosis is roughly half the corresponding value in deep waters, which is already negligible as the sea state is broadbanded.
Clearly, the Gaussian adjustment time $t_{c}\sim O(T_{0})\sim15$ seconds,
indicating that nonlinear wave-wave interactions are negligible. Indeed,
$C_{4}^{d}$ is slightly negative, implying a defocusing wave regime
due to the short-crestedness of the sea state whereas the non-zero
and positive $C_{4}^{b}$ component indicates that bound nonlinearities are not negligible. %
\begin{comment}
Note that even by artificially doubling the spectral or angular spreading,
dynamic kurtosis values stay below $O(10^{-1}-10^{-2})$. 
\end{comment}
 
Clearly, to date numerical simulations of the Euler equations are computationally expensive and unfeasible for an operational forecasting of extreme waves. One should consider  Janssen's general third-order expression for kurtosis, which is valid for any depth and arbitrary spectra. This is a complex formula whose implementation requires care and effort beyond the scope of this paper.

In this work we propose that statistical predictions of extreme waves of realistic oceanic seas can be based on Adler-Taylor's (2009) theory of random fields coupled with the Tayfun (1980) and Janssen (2009) models to account for both second and third-order nonlinearities. In the following, we will first present the theory of space-time extremes for Gaussian seas (\cite{Fedele2012}). Adler-Taylor theory is then extended to third-order non-Gaussian seas and apply it to study the statistical properties of the Andrea rogue wave event.

\section{Space-time (ST) extremes}

In accord with ERA-interim reanalysis, we can assume that sea states are both stationary in a time interval $D\leq D_{\mathrm{ERA}}=6$ hours and homogeneous over an area $A\leq A_{\mathrm{ERA}}=100^{2}$ $\mathrm{km^{2}}$. Then, the free surface $\eta(\mathbf{x},t)$ can be modeled
as a three-dimensional (3-D) homogeneous Gaussian random field over
the space-time volume $\mathrm{\Omega}$ defined by the area $A$
and the time interval $D$, and $\mathbf{x}=(x,y)$ denotes the horizontal coordinate vector. 
Thus, the associated probability distributions at any point
of the volume is the same and Gaussian. Drawing on \cite{adler1981geometry}, we next consider the Euler characteristics (EC) of excursion sets of $\eta$
defined as follows. Given a threshold $z$, the excursion set $U_{\mathit{\mathrm{\Omega}}}(z)$
is the part of $\mathrm{\Omega}$ within which $\eta$ is above~$z$:
\begin{equation}
U_{\mathit{\mathrm{\text{\ensuremath{\Omega}}}}}(z)=\{(\mathbf{x},t)\in\text{\ensuremath{\Omega}}:\eta(\mathbf{x},t)>z\}.\label{exc}
\end{equation}
In 1-D Gaussian processes, the EC simply counts the number of $z$-upcrossings. Thus, (\ref{exc}) provides the generalization of this concept to to higher dimensions. Indeed,
for two dimensional (2-D) random fields, EC counts the number
of connected components minus the number of holes of the respective
excursion set. In 3-D sets instead, EC counts the number of connected
volumetric components of the set, minus the number of holes that pass
through it, plus the number of hollows inside. Further, the probability
of exceedance that the global maximum of $\eta$ over $\mathrm{\Omega}$,
say $\eta_{\max}$, exceeds $z$ depends on the domain size and it
is well approximated by the expected EC of the excursion set, provided
that $z$ is sufficiently high (\cite{adler1981geometry,adler2000excursion,adler2009random}).
Intuitively, as $z$ increases the holes and hollows
in the excursion set $U_{\mathrm{\Omega}}(z)$ disappear until each
of its connected components includes just one local maximum of $\eta$,
and EC counts the number of local maxima. For very large thresholds, EC equals 1 if the global maximum exceeds the threshold and 0
otherwise. Thus, EC of large excursion sets is a binary random
variable with states 0 and 1, and, for large $z$, 
\begin{equation}
\mathbf{\mathrm{Pr}}\left\{ \eta_{\max}>z\right\} =\mathrm{Pr}\left\{ EC\left(U_{\mathrm{\Omega}}(z)\right)=1\right\} =\left\langle EC\left(U_{\Omega}(z)\right)\right\rangle ,\label{PM}
\end{equation}
where angled brackets denote expectation. This heuristic identity
has been proved rigorously to hold up to an error that is in general
exponentially smaller than the expected EC approximation (\cite{adler2009random,adler2000excursion}).
For 3-D random fields, which are of interest in oceanic applications,
the probability $P_{\mathrm{ST}}(\xi;A,D)$ that the maximum surface
elevation $\eta_{\max}$ over the area $A$ and during a time
interval $D$ exceeds the threshold $\xi H_{s}$ is given by (\cite{adler2009random})
\begin{equation}
\begin{split}
\mathrm{P_{ST}}(\xi;A,D)
&=\mathrm{Pr}\left\{ \eta_{\max}>\xi H_{s}\right\}\\ &=(16M_{3}\xi^{2}+4M_{2}\xi+M_{1})\mathrm{P_{\mathrm{R}}(\xi)},\label{PV}
\end{split}
\end{equation}
where
\begin{equation}
P_{\mathrm{R}}(\xi)=\mathrm{Pr}\left\{ h>\xi H_{s}\right\} =\exp(-8\xi^{2})\label{Ra}
\end{equation}
is the Rayleigh exceedance probability of the crest height $h$ of
a time wave observed at a single point within $A$.
Here, $M_{1}$ and $M_{2}$ are the average number of 1-D and 
2-D waves that can occur on the edges and boundaries of the volume $\mathrm{\Omega}$, and $M_{3}$ is the average number of 3-D waves that can occur within the volume (\cite{Fedele2012}). These all depend on the directional wave spectrum
and are given in Appendix~A. 
%Finally, it is noted that Piterbarg (1995)\nocite{piterbarg1995}
%also derived an asymptotic expansion of the probability in (\ref{PM})
%for large higher dimensional Gaussian maxima via generalized Rice
%(1944,1945) formulas\nocite{Rice1944,Rice1945}.

A statistical indicator of the geometry of space-time extremes in
the volume $\Omega$ is the wave dimension $\beta$ defined by \cite{Fedele2012} as
\begin{equation}
\beta=3-\frac{4M_{2}\xi_{\mathrm{m}}+2M_{1}}{16M_{3}\xi_{\mathrm{m}}^{2}+4M_{2}\xi_{\mathrm{m}}+M_{1}},\label{beta}
\end{equation}
where $\xi_{\mathrm{m}}$ is the most probable surface elevation, or modal value which,
according to Gumbel (1958) and Eq. (\ref{PV}), satisfies 
\begin{equation}
P_{\mathrm{ST}}(\xi_{\mathrm{m}};A,D)=(16M_{3}\xi^{2}_{\mathrm{m}}+4M_{2}\xi_{\mathrm{m}}+M_{1})\mathrm{\mathit{P}_{\mathrm{R}}(\xi_{\mathrm{m}})}=1.\label{xm0}
\end{equation}
Drawing on \cite{Fedele2012}, the correct average number of space-time waves of dimension $\beta$ occurring within the space-time volume $\mathrm{\Omega}$ spanned by area $A$ and time interval $D$ is 
\begin{equation}
N_{\mathrm{ST}}(A,D)=\frac{16M_{3}\xi^{2}_{\mathrm{m}}+4M_{2}\xi_{\mathrm{m}}+M_{1}}{(4\xi_{\mathrm{m}})^{\beta-1}}.\label{NST}
\end{equation}
The parameter $\beta$ represents a scale dimension of waves, i.e. the relative scale of a space-time
wave with respect to the volume\textquoteright s size and $1\leq\beta\leq3$. In particular,
if wave extremes are 3-D ($\beta>2$) they are expected to occur within
the volume $\Omega$ away from the boundaries, whereas the limiting
case of 1-D time extremes ($\beta\sim1$) occur for time waves observed
at a single point. Furthermore, \cite{Fedele2012} showed that space-time
extremes are larger than time extremes in agreement with
recent stereo measurements of oceanic sea states (\cite{fedele2013}, see also Fig.~\ref{FIG7}).

The bottom-left panel of Fig.~\ref{FIG4} shows the map of the estimated wave dimension $\beta$ for the North Sea area at the peak time of the Andrea storm for sea states of duration $D=3$ hours over the cell area $A_{\mathrm{ERA}}$. Clearly, sea states were short-crested and extremes are roughly 3-D, indicating
that the area considered is large compared to the mean wavelength. Thus, in accord
with Boccotti's (2000)\nocite{Boccotti2000} QD theory, a space-time extreme most likely coincides with the crest of a focusing wave group that passes through the area as described below.

\section{Extreme nonlinear wave groups}

Drawing on Boccotti's (2000) QD theory, \cite{Fedele2009} showed that for second-order weakly nonlinear waves the expected space-time
dynamics near a large wave crest is that of a stochastic wave group whose free surface is described by
\begin{equation}
\zeta_{c}/H_s=\xi_{0}\zeta_{1}+\xi_{0}^{2}\zeta_{2},
\end{equation}
where $\xi_{0}=h_{0}/H_{s}$ is the dimensionless linear crest height, 
\begin{equation}
\zeta_{1}(\mathbf{X},T)=\Psi(\mathbf{X},T)
\end{equation}
is the linear component, 
\begin{equation}\label{z1}
\begin{split}
\Psi(\mathbf{X},T)&=\frac{\left\langle \eta(\mathbf{x},t)\eta(\mathbf{x}+\mathbf{X},t+T)\right\rangle }{\sigma^{2}}\\
&=\int\frac{S_{1}}{\sigma^{2}}\cos\left(\chi_{1}\right)d\omega_{1}d\theta_{1}
\end{split}
\end{equation}
is the space-time covariance of $\eta$ (\cite{Boccotti2000}) and
\begin{equation}
\begin{split}
\zeta_{2}=&\int\frac{S_{1}S_{2}}{\sigma^{3}}\Bigl(A_{12}^{+}\cos\left(\chi_{1}+\chi_{2}\right)+\\
&\qquad\qquad A_{12}^{-}\cos\left(\chi_{1}-\chi_{2}\right)\Bigr)
d\omega_{1}d\theta_{1}d\omega_{2}d\theta_{2}\label{z2}
\end{split}
\end{equation}
is the second-order correction. Here, $S_{j}=S(\omega_{j},\theta_{j})$
and $\chi_{j}=\mathbf{k_{\mathrm{j}}}\cdot\mathbf{X}-\omega_{j}T$,
where $\mathbf{X=\mathit{\left(X,Y\right)}}$ and $\mathbf{k_{\mathrm{j}}=\mathit{\left(k_{j}\sin\theta_{j},k_{j}\cos\theta_{j}\right)}}$
with $k_{j}\mathrm{tanh}(k_{j}d)=\omega_{j}^{2}/g$ from linear dispersion,
and the coefficients $A_{12}^{\ensuremath{\pm}}$ can be found in
\cite{sharma1979development}.
\begin{comment}
Note that this solution is valid for generic sea states including crossing seas. 
\end{comment}
 For generic sea states, the largest nonlinear crest amplitude is attained
at the focusing point ($\mathbf{X}=\mathbf{0},T=0$) and given by
\begin{equation}
\xi=\xi_{0}+2\mu\xi_{0}^{2},\label{csi}
\end{equation}
where $\xi=h/H_{s}$ is the nonlinear crest height. The Tayfun wave steepness $\mu=C_3/3$
relates to the wave skewness $C_{3}$ of surface elevations. 
For oceanic applications in deep waters, \cite{Fedele2009} proposed
the approximation
\begin{equation}
\mu\sim\mu_{a}=\mu_{m}\left(1-\nu+\nu^{2}\right),\label{mua}
\end{equation}
where $\mu_{m}=k_{m}\sigma$, which is an upper bound for $C_3$.  From the linear dispersion relation
$k_{m}=\omega_{m}^{2}/g$ is the wavenumber corresponding to the mean
spectral frequency $\omega_{m}=m_{001}/m_{000}$,
\begin{equation}
 \nu=\sqrt{m_{000}m_{002}/m_{001}^{2}-1}\label{nu}
 \end{equation}
is the spectral bandwidth and $m_{ijk}$ are spectral moments (see Appendix~B).  In intermediate water depths, the narrowband approximations lead to \citep{Tayfun2006} 
\begin{equation}
\mu\sim\mu_{S}=\mu_{m}f_{S},\label{mua-1}
\end{equation}
where $f_{S}=D_{1}+D_{2}$, with
\begin{align*}
&D_{1}=\frac{1}{2}\frac{4n-1}{n^{2}\mathrm{tanh}(q_{m})-q_{m}},\\ &D_{2}=\frac{\mathrm{cosh(}q_{m})\left[2+\mathrm{cosh}(2q_{m})\right]}{2\mathrm{sinh^{3}(}q_{m})},
\end{align*}
$n=\left[1+2q_{m}/\mathrm{sinh(2}q_{m})\right]/2$, $q_{m}=k_{m}d$ and $d$ is the water depth. The coefficients $D_{1}$ and $D_{2}$
arise from the frequency-difference and frequency-sum terms, i.e.
$A_{12}^{-}$ and $A_{12}^{+}$, in Eq.~\eqref{z2} as the spectral
bandwidth $\nu\rightarrow0$. A general expression for skewness valid for finite depths and arbitrary spectra is given by \cite{JanssenJFM2009}. The NB limit of Janssen's skewness yields the same Eq.~\eqref{mua-1} derived by~\cite{Tayfun2006}.  

Further, the wave trough following the large crest occurs at $t=T^{*}$, where $T^{*}$ is the abscissa of
the first minimum of the time covariance $\psi(T)=\Psi(\mathbf{X}=\mathbf{0},T)$ \citep{Boccotti2000}. 
Second-order nonlinearities do not affect  the  crest-to-trough heights of large waves since wave crests and  troughs  are displaced upward equally.
Thus, the maximum second-order nonlinear crest-to-trough height observed at a point in time remains essentially the same as that of the linear
group $\zeta_{1}$ and given by $H/H_s=\xi_{0}\left(1+\psi^{*}\right)$, 
where $\psi^{*}=\psi(T^{*})$ is the Boccotti's (2000)\nocite{Boccotti2000}
narrowbandedness parameter. Note that for narrowband waves $\psi^{*}\rightarrow1$. 
The left panels of Fig.~\ref{FIG4} show the maps of the 
Tayfun NB steepness $\mu$ (top) from Eq. (\ref{mua-1}) and Boccotti $\psi^{*}$ (bottom) estimated
from the hindcasted ERA-interim sea state in which the Andrea wave occurred. It appears that $\psi^{*}\sim0.75$ as the characteristic value of short-crested sea states dominated by wind waves (\cite{Boccotti2000}).
%and that $\mu\sim0.12$ as the maximum wave steepness typical of oceanic storms \citep{Tayfun2008}. %
At the Ekofisk location, the NB prediction gives $\mu=C_3/3\sim0.05$ in fair agreement with the predicted values from numerical simulations of the Euler equations \citep{Bitner_Andrea2014,Dias2015} and actual laser measurements \citep{Magnusson2013}).

We have shown that that third-order quasi-resonant interactions do not play any role in the formation of large waves in realistic oceanic seas. However, the effects of third-order bound nonlinearities on the prediction of maximum crest and wave heights must be accounted for as addressed in the next section. 

\section{A new stochastic third-order space-time (FST) model} 

Drawing on \cite{Fedele2008a,Fedele2012} and \cite{TayfunFedele2007}, we propose
a new stochastic model, hereafter referred to as $FST$, which accounts for both second and third-order nonlinearities in the prediction of space-time extremes. In particular, consider a 3-D non-Gaussian field over an area $A$ for a time period of $D$. Clearly, the area cannot be too large since the wave field may not be homogeneous. The duration should be short so that
spectral changes occurring in time are not so significant and the sea state can be assumed as stationary.
Then, the third-order nonlinear probability $P_{\mathrm{FST}}^{(nl)}(\xi;A,D)$ that the maximum surface elevation $\eta_{\max}$ over the area $A$ and during
the time interval $D$ exceeds the generic threshold $\xi H_{s}$ is equal to the probability of exceeding the threshold $\xi_{0}$, which accounts for kurtosis effects only, 
that is 
\begin{equation}
P_{\mathrm{FST}}^{(nl)}(\xi;A,D)=P_{\mathrm{ST}}(\xi_{0};A,D)\left(1+\Lambda\xi_{0}^{2}(4\xi_{0}^{2}-1)\right).\label{Pid}
\end{equation}
The Gaussian probability of exceedance $P_{\mathrm{ST}}$ is given in Eq. (\ref{PV}) and the amplitude $\xi$, which accounts for both skewness and kurtosis effects, relates to $\xi_0$ via the Tayfun (1980) quadratic equation 
\begin{equation}
\xi=\xi_{0}+2\mu\xi_{0}^{2}.\label{sub2}
\end{equation}
Further, the parameter
\begin{equation}
\Lambda=\lambda_{40}+2\lambda_{22}+\lambda_{04}\label{la}
\end{equation}
is a measure of third-order nonlinearities and it is a function of the fourth order cumulants $\lambda_{nm}$ of the wave surface $\eta$ and its Hilbert transform $\hat{\eta}$ \citep{TayfunFedele2007}
\begin{equation}
\lambda_{nm}=\left\langle \eta^{n} \hat{\eta}^m\right\rangle/\sigma^{n+m}+(-1)^{m/2}(n-1)(m-1),
\end{equation}
where $\sigma$ is the standard deviation of the wave surface and $n+m=4$. Drawing on \cite{Janssen2006}, we assume the following relations between cumulants
\begin{equation}
\lambda_{22}=\lambda_{40}/3,\quad\lambda_{04}=\lambda_{40},\label{cumul}
\end{equation}
and Eq.~\eqref{la} simplifies to
\begin{equation}
\Lambda=\Lambda_{\mathrm{appr}}=\frac{8\lambda_{40}}{3},\label{Lambda}
\end{equation}
which will be used in this work. Then, Eq.~\eqref{Pid} reduces to a modified Edgeworth-Rayleigh (MER) distribution \citep{Janssen2006}. For realistic oceanic seas the kurtosis $\lambda_{40}\sim\lambda_{40}^b$ is only due to bound nonlinearities. To date, the validity of the cumulant relations in Eq.~\eqref{cumul} has been proven to hold for second-order NB waves only \citep{TayfunLo1990}. Further studies are desirable to determine the general expressions of fourth order cumulants of the wave surface relying on Janssen's (2009) third-order Hamiltonian formulation, but this is beyond the scope of this paper. 

Given the probability structure of the wave surface defined by Eq.~\eqref{Pid}, the nonlinear mean maximum surface or crest height $\overline{h}_{\mathrm{FST}}=\xi_{\mathrm{FST}}H_s$ attained over the area $A$ during a time interval $D$ is given, according to Gumbel (1958), by
\begin{equation}
\begin{split}
\xi_{\mathrm{FST}}&=\frac{\overline{h}_{\mathrm{FST}}}{H_{s}}\\
&=\xi_{\mathrm{m}}+2\mu\xi_{\mathrm{m}}^{2}+\frac{\gamma_{e}\left(1+4\mu\xi_{\mathrm{m}}\right)}{16\xi_{\mathrm{m}}-\frac{32M_{3}\xi_{\mathrm{m}}+4M_{2}}{16M_{3}\xi_{\mathrm{m}}^{2}+4M_{2}\xi_{\mathrm{m}}+M_{1}}-\Lambda G(\xi_{\mathrm{m}})  },\end{split}
\label{xist}
\end{equation}
where the most probable surface elevation value $\xi_{\mathrm{m}}$ satisfies $P_{\mathrm{FST}}(\xi_{\mathrm{m}};A,D)=1$, or equivalently from Eq.~\eqref{Pid}
\[
(16M_{3}\xi_{\mathrm{m}}^{2}+4M_{2}\xi_{\mathrm{m}}+M_{1})\mathrm{\mathit{P}_{\mathrm{R}}(\xi_{\mathrm{m}})}\left(1+\Lambda\xi_{\mathrm{m}}^{2}(4\xi_{\mathrm{m}}^{2}-1)\right)=1,
\]
and
\[
G(\xi_{\mathrm{m}})=\frac{2\xi_{\mathrm{m}}(8\xi_{\mathrm{m}}^{2}-1)}{1+\Lambda\xi_{\mathrm{m}}^{2}(4\xi_{m}^{2}-1)}.\label{gm}
\]
The nonlinear mean maximum surface or crest height $h_{\mathrm{T}}$ expected at a point during the time interval $D$ follows from Eq.~\eqref{xist} by setting $M_2=M_3=0$ and $M_1=N_{\mathrm{D}}$, namely 
\begin{equation}
\xi_{\mathrm{T}}=h_{\mathrm{T}}/H_{s}=\xi_{\mathrm{m}}+2\mu\xi_{\mathrm{m}}^{2}+\frac{\gamma_{e}\left(1+4\mu\xi_{\mathrm{m}}\right)}{16\xi_{\mathrm{m}}-\Lambda G(\xi_{\mathrm{m}}) },\label{ht}
\end{equation}
where, now, $\xi_{\mathrm{m}}$ satisfies
\[
N_{\mathrm{D}}P_{\mathrm{R}}(\xi_{\mathrm{m}})\left(1+\Lambda\xi_{\mathrm{m}}^{2}(4\xi_{\mathrm{m}}^{2}-1)\right)=1.
\]
Here, $N_{\mathrm{D}}=D/\bar{T}$ denotes the number of wave occurring during $D$ and $\bar{T}$ is the mean up-crossing period (see Appendix~B). The linear mean counterpart follows from Eq.~\eqref{ht} by setting $\mu=0$ and $\Lambda=0$.

Drawing on the Boccotti (2000) distribution for wave heights, the third-order nonlinear mean maximum wave height expected at a point is given by
\begin{equation}
 \overline{H}_{\mathrm{T}}=\overline{h}_{\mathrm{T}}^{(\Lambda)}\sqrt{\ensuremath{2\left(1+\psi^{*}\right)}},\label{Hmax}
 \end{equation}
where $\overline{h}_{\mathrm{T}}^{(\Lambda)}$ is the nonlinear mean maximum crest height  that accounts for kurtosis effects only. This follows from Eq.~\eqref{ht} by setting $\mu=0$, namely
\begin{equation}
\overline{h}_{\mathrm{T}}^{(\Lambda)}/H_{s}=\xi_{\mathrm{m}}+\frac{\gamma_{e}}{16\xi_{\mathrm{m}}-\Lambda G(\xi_{\mathrm{m}}) },\label{htf}
\end{equation}
where
\[
N_{\mathrm{D}}P_{\mathrm{R}}(\xi_{\mathrm{m}})\left(1+\Lambda\xi_{\mathrm{m}}^{2}(4\xi_{\mathrm{m}}^{2}-1)\right)=1.
\]
For narrowband waves, $\psi^{*}=1$ and Eq.~\eqref{Hmax} reduces to the MER model proposed by \cite{Janssen2006}. Clearly, this overestimates wave heights in short-crested or multidirectional seas as, in general, $\psi^{*}<1$. 

When the lateral dimension $\ell=\sqrt{A}$ is much larger than the average
wavelength $L_{0}$, the maximum surface height occurs most likely within the area of interest and not on its boundaries. In this case, on average the number of 3-D waves is much larger than the numbers of 1-D and 2-D waves that can occur on the boundaries, i.e. $M_3\gg{M_{2}}$ and $M_{1}$. Keeping only the leading term in Eq. (\ref{xist}) yields the following asymptotic value of the expected surface wave height maximum over large areas
\begin{equation}
\xi_{\mathrm{FST}}^{(3D)}=\frac{h_{\mathrm{FST}}^{(3D)}}{H_s} =\xi_{\mathrm{m}}+2\mu\xi_{\mathrm{m}}^{2}+\frac{\gamma_{e}\left(1+4\mu\xi_{\mathrm{m}}\right)}{16\xi_{\mathrm{m}}-\frac{2}{\xi_{\mathrm{m}}}-\Lambda G(\xi_{\mathrm{m}}) },\label{xi3d}
\end{equation}
where
\[ 16M_{3}\xi_{\mathrm{m}}^{2}\mathrm{\mathit{P}_{\mathrm{R}}(\xi_{\mathrm{m}})}\left(1+\Lambda\xi_{\mathrm{m}}^{2}(4\xi_{\mathrm{m}}^{2}-1)\right)=1.
\]
For 2-D waves only ($M_{1}=M_{3}=0$ in Eq. (\ref{xist})) 
\begin{equation}
\xi_{\mathrm{FST}}^{(2D)}=\frac{\xi_{\mathrm{FST}}^{(2D)}}{H_s} =\xi_{\mathrm{m}}+2\mu\xi_{\mathrm{m}}^{2}+\frac{\gamma_{e}\left(1+4\mu\xi_{\mathrm{m}}\right)}{16\xi_{\mathrm{m}}-\frac{1}{\xi_{\mathrm{m}}}-\Lambda G(\xi_{\mathrm{m}}) },\label{xist-1}
\end{equation}
where
\[ 4M_{2}\xi_{\mathrm{m}}\mathrm{\mathit{P}_{\mathrm{R}}(\xi_{\mathrm{m}})}\left(1+\Lambda\xi_{\mathrm{m}}^{2}(4\xi_{\mathrm{m}}^{2}-1)\right)=1.
\]
If one ignores kurtosis effects ($\Lambda=0$), Eqs.~\eqref{xi3d} and \eqref{xist-1} reduce to the 2-D and 3-D analogues of the Piterbarg-Tayfun (PT) model formulated by \cite{Juglard2005}, and hereafter referred to as 2D-PT  and 3D-PT respectively (see also \cite{piterbarg1995} and~\cite{forristall2011}). 
%As wave breaking is neglected in both ATT and PT models, these are expected to give conservative %estimates of wave maxima.  
%Hereafter, we focus on the short-term or sea state analysis of the Andrea event ($D=3$ hours). 

In offshore applications, the interest is in the expected wave maxima
over small areas such as those covered by oil rigs, i.e. $\ell\leq L_{0}$.
In this range, \cite{fedele2013} have shown that the boundary corrections accounted by both terms $M_{1}$ and $M_{2}$ are important for a correct estimation of Euler characteristics and expected maxima (see also \cite{forristall2015}). Hence, they cannot be ignored as has been assumed by \cite{RomoloArena2015}, since maximum surface heights expected over small areas are underestimated. This point is elaborated further in the next section and demonstrated explicitly by way of the results displayed in Fig.~\ref{FIG7}.

\begin{figure}[t]
\centering\includegraphics[width=1.06\columnwidth]{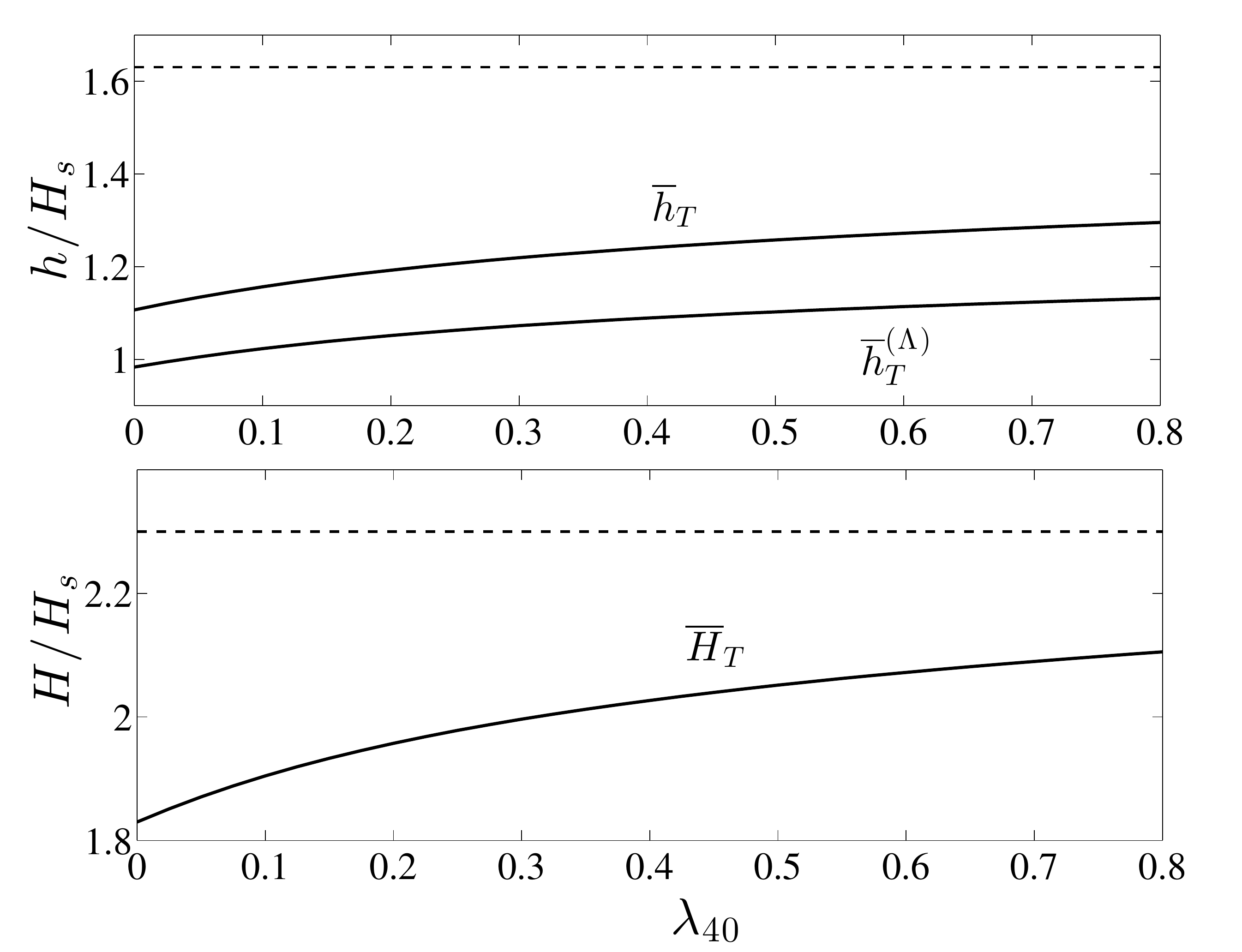} 
\caption{Andrea wave: (top panel) nonlinear mean maximum surface or crest height $\overline{h}_{T}$ expected at a point as a function of excess kurtosis $\lambda_{40}$, which accounts for both second-order (skewness) and third-order (kurtosis) nonlinearites. The prediction $\overline{h}_{T}^{(\Lambda)}$ of average heights that accounts for third-order kurtosis effects only is also shown. (Bottom panel) nonlinear mean maximum wave height expected at a point ($\overline{H}_{T}$ ) as a function of excess kurtosis $\lambda_{40}$. Dashed lines denote the actual Andrea crest and wave height values ($h_{osb}=1.63H_s$ and $H_{obs}=2.3H_s$).} 
\label{FIG6} 
\end{figure}

\begin{figure*}[t]
\centering\includegraphics[width=1.05\textwidth]{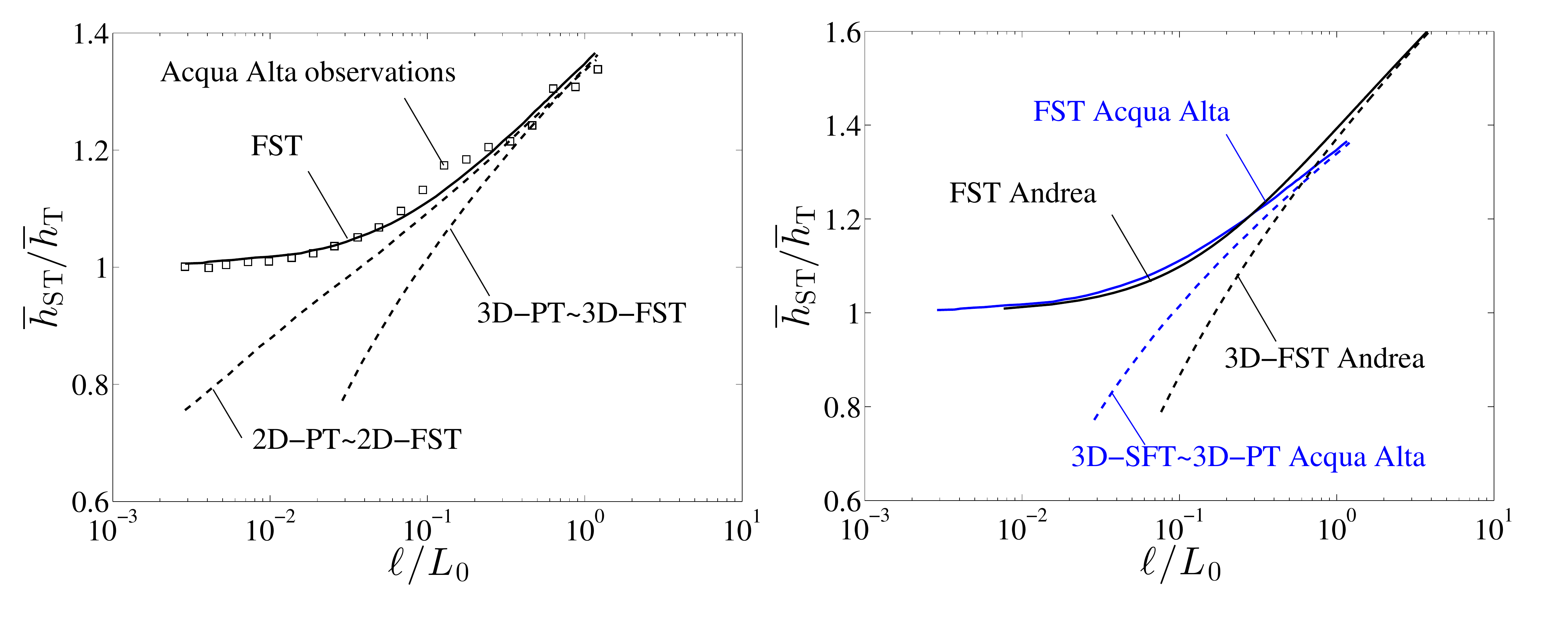}
\caption{Left panel, Acqua Alta stereo measurements (\cite{fedele2013}): comparison between the observed ratio of space-time and time maximum surface heights (hollow squares) and theoretical averages $\overline{h}_{\mathrm{FST}}/\overline{h}_{\mathrm{T}}$ as a function of the lateral side length $\ell$ normalized to the average wavelength $L_{0}$. Fedele space-time (FST) model (solid line) and 2D and 3D Piterbarg-Tayfun (PT)  models nearly the same as 2D- and 3D-FST (dashed lines). Right panel, Andrea ERA-interim predictions: expected ratio of mean areal and point maximum surface heights estimated at the Ekofisk location (FST, black line; 3D-FST, dashed line). For comparison the estimated ratios for Acqua Alta are also shown (FST, blue solid line; 3D-FST~2D-PT, blue dashed line).}
\label{FIG7} 
\end{figure*}

As an application, consider now the hindcasted sea state during which the Andrea wave occurred (wave steepness $\mu=C_3/3\sim0.05$  and excess kurtosis $\lambda_{40}=3C_4\sim0.1$ from \cite{Dias2015}).  The top panel of Fig.~\ref{FIG6} displays the predictions from Eq. (\ref{ht}) for the third-order mean maximum surface or crest height $\overline{h}_{\mathrm{T}}$ expected at a point as a function of the excess kurtosis $\lambda_{40}$ for a typical $D=3$-hour sea state. From Eq. \ref{htf}, the mean height $\overline{h}_{\mathrm{T}}^{(\Lambda)}$ that accounts for kurtosis effects only is also shown. Comparing the two predictions, it is clear that second-order nonlinearities cannot be neglected as they yield a substantial $15\%$ increase in crest height in comparison to the modest $5\%$ increase due to kurtosis ($\sim0.1$) from the linear estimates ($\lambda_{40}=0$). Note that even for unrealistic values of kurtosis ($\sim0.8$) the predictions are well below the observed actual Andrea crest height $h_{obs}=1.63H_s$ (dashed line). Further, the  bottom panel of  Fig.~\ref{FIG6} displays the mean maximum wave height $\overline{H}_{\mathrm{T}}$ expected at a point from Eq. (\ref{Hmax}). This is smaller than the observed actual Andrea wave height $H_{obs}=2.3H_s$. 

In summary, averages of maxima do not explain the actual large Andrea time crest and wave heights observed at the measurement site. This point will be discussed and addressed later in section \ref{hqsection}. 

\section{ ST analysis of the Andrea wave}

In this section we will study the space-time properties of the Andrea storm in comparison to 
stereo wave measurements at the Acqua Alta site of a short-crested sea state dominated by bora winds  (experiment 2, see \cite{fedele2013}). 
For example, Fig.~\ref{FIG7} shows the prediction of space-time
extremes at the Acqua Alta site and from the hindcasted ERA-interm Andrea sea state during which the rogue wave occurred. In particular, the left panel of the figure
shows the Acqua Alta observed ratio of space-time and time maximum surface heights over a period of $D\sim0.5$ hours as a function of the lateral dimension $\ell$ normalized to the average
wavelength $L_{0}\sim19$ m (hollow squares, maximum length $\ell_{max}\sim24$ m). 
%This is given by the geometric mean $\sqrt{L_{x}L_{y}}$
%of the average wavelenghts along $x$ and $y$ ($L_{0}\sim19$ m).
In the same panel, we compare the theoretical $FST$ ratio $\overline{h}_{\mathrm{FST}}/\overline{h}_{\mathrm{T}}$,  which accounts for boundary corrections associated with $M_1$
and $M_2$ in Eq. (\ref{xist}), and the asymptotic 3D-FST from Eq. (\ref{xi3d}), which is valid over large areas and ignores boundary effects. The 2D-FST model from Eq. (\ref{xist-1}) is also shown. They both nearly coincide with 2D-PT and 3D-PT as the measured kurtosis at Acqua Alta was almost Gaussian. 
As one can see, the boundary contributions cannot be neglected over areas with lateral dimension comparable to or smaller than the typical wavelength. Indeed, both 2D- and 3D-FST underestimate the observed ratios when $\ell\leq L_{0}$. Note that for $\ell\sim L_{0}$ the
observed space-time wave surface maximum $\overline{h}_{\mathrm{FST}}$ is 1.4 times
larger than the time maximum $\overline{h}_{\mathrm{T}}$ at a single point.

A similar trend is also observed for the expected space-time extremes
of the Andrea sea state for the same time period of $D=0.5$ hours. Specifically, 
the right panel of Fig.~\ref{FIG7} displays the mean maximum surface height ratios $\overline{h}_{\mathrm{FST}}/\overline{h}_{\mathrm{T}}$ (FST, black curve)  as a function of $\ell/L_{0}$ estimated
at the Ekofisk location, where the ERA-interim predicts a mean wavelength $L_{0}\sim150$ m. 
% and at the grid points (3D-ATT, red curves) where $H_{s}>4$ m ($L_0=60-160$ m). 
%The estimated curves are very close to each other, indicating that the sea states are multidirectional, %or short-crested, and have similar spectral characteristics. This is reflected in the small range of %variability of the wave dimension $\beta$, which varies within the interval $2.7-2.8$, as shown in the %left-bottom panel of Fig. (\ref{FIG3F}).
%that the parameter $R$ estimated from Eq. (\ref{RR})
%is within the small range of $10-20$. Further, the right panel shows that
Note that the FST  ratios for Acqua Alta and Andrea (blue and black solid curves) are close to each other for $\ell\leq L_0$. These results are very encouraging as the observed sea state at Acqua Alta was in deep waters (measured $H_s=1.09$ m and $d/L_0=1.25$, with $d=16$ m) whereas the Andrea sea state was in intermediate waters (ERA-interim estimates $H_s=8.3$ m and $d/L_0=0.49$, with $d=74$ m). Although ERA-interim underestimates the actual value of $H_s$ at Ekofisk ($=9.2$ m), it appears that the maximum surface height ratio $\overline{h}_{\mathrm{FST}}/\overline{h}_{\mathrm{T}}$ is slightly sensitive to the significant wave height level and just depends on average spectral properties of the sea state. Further studies are desirable to investigate possible statistical similarities and universal laws for space-time extremes in wind sea states, but this is beyond the scope of this paper.

Clearly, space-time extremes cannot explain the actual large Andrea time crest height observed at the measurement point. The space-time analysis of the Andrea storm simply predicts that the maximum surface wave height over the platform footprint area
is $\sim20\ensuremath{\%}$ higher than the average maximum surface height that is expected at a fixed point within the same area ($\ell/L_0\sim0.3$, see right panel of Fig.~\ref{FIG7}).
Indeed, in relatively short-crested directional
seas such as those observed around the Ekofisk area, it is very unlikely
that the observed crest actually coincides with the largest crest
of a group of waves propagating in space-time. In contrast, in accord
with Boccotti's (2000) QD theory, it is most likely
that the sea surface was in fact much higher somewhere near the measurement point. 
%In the following we formalize the interpretation of rogue wave %observations in \cite{Janssen2009}, who proposed to adopt the %notion that the measured large crest is just a random draw from %a crest population that refers to the tail of the associated %probability distribution. 

\section{Threshold $h_q$ exceeded with probability $q$ by the maximum surface height}\label{hqsection}

Consider the average duration of $D=3$ hours for the Andrea sea state in which the rogue wave was observed, considering that the ERA time resolution is of $D_{ERA}=6$ hours.  Monte Carlo simulations yield wave steepness $\mu\sim0.05$ and excess kurtosis $\lambda_{40}\sim0.1$ \citep{Bitner_Andrea2014,Dias2015}. %The difference in the predicted extreme values is less than $3\%$ for the two durations. 
It follows that the mean maximum nonlinear wave surface height $\overline{h}_\mathrm{T}$ expected at a point is $\sim1.2H_s$ (see Fig. (\ref{FIG6}). 
%(see top-left panel of Fig. (\ref{FIG6F})). 
This is lower than the actual value $h_{obs}\sim1.63H_s$ observed. Statistically, this means that in an ensemble of $M$ realizations of $3$-hour sea states, each of which has similar statistical structure to the Andrea wave field, all the maximum surface heights observed at a point will exceed $1.2H_s$. Clearly, the maximum surface height at a point can reach or exceed the actual observed value $1.63H_s$ only in few realizations out of the ensemble of $M$ sea states.

To characterize such rare occurrences in weakly nonlinear third-order random seas one can consider the threshold $h_q=\xi_q H_s$ exceeded with probability $0\le q \le1$ by the maximum surface height $\eta_{\max}$ over an area $A$ during a sea state of duration $D$. This satisfies (see Appendix~C)
\begin{equation} 
P_{\mathrm{FST}}^{(nl)}(\xi_q;A,D)=q,\label{PNL1}
\end{equation} 
where $\xi_q=\xi_{0,q}+2\mu\xi_{0,q}^{2}$, $\xi_{0,q}$ follows from Eq. (\ref{EE}) and the nonlinear probability of exceedance $P_{\mathrm{FST}}^{(nl)}$ is given in Eq. (\ref{Pid}). Further, the conditional mean $\overline{h}_q=\left\langle\eta_{\max}|\eta_{\max}>h_{q}\right\rangle$ is given by 
\begin{equation}
\overline{h}_{q}=h_{q}+(\chi_{1,q}+\mu\chi_{2,q}+\Lambda\chi_{3,q}+\mu\Lambda\chi_{4,q})H_s/q,\label{h1n}
\end{equation}
where the coefficients $\chi_{j,q}$ depend on $q$ as well as on the spectral parameters $M_{j}$ and $\xi_{q}$.
For Gaussian waves, the linear threshold $h_{q}^{(L)}$ and conditional mean $\overline{h}_{q}^{(L)}$ follow by setting both $\mu=0$ and $\Lambda=0$ in Eqs. (\ref{PNL1}) and (\ref{h1n}) respectively. For second-order waves the associated threshold $h_{q}^{(\mu)}$ and mean $\overline{h}_{q}^{(\mu)}$ follow by setting $\Lambda=0$ only.  
As $q$ becomes smaller the threshold $h_{q}$ tends to coincide with the mean $\overline{h}_{q}$ since terms $\chi_{j,q}/q$ in Eq. (\ref{h1n}) tend to zero, and similarly for linear and second-order thresholds. %For ERA-interim Andrea data the difference between the two quantities is less than $5\%$ ($3\%$) for $q<1/10$ ($1/100$).

In this work, we consider the time thresholds $h_{\mathrm{T},q}^{(L)}$, $h_{\mathrm{T},q}^{(\mu)}$ and $h_{\mathrm{T},q}$ at a point and conditional averages $\overline{h}_{\mathrm{T},q}^{(L)}$, $\overline{h}_{\mathrm{T},q}^{(\mu)}$ and $\overline{h}_{\mathrm{T},q}$. These can be computed using $M_{2}=M_{3}=0$ and $M_1=N_{\mathrm{D}}$ in Eqs. (\ref{PNL1}) and (\ref{h1n}). Here, $N_\mathrm{D}$ denotes the number of zero-crossing waves occurring during the time interval $D$. 
Further, note that $\overline{h}_{\mathrm{T},q=1}$ coincides with the mean maximum third-order surface height  $\overline{h}_{\mathrm{T}}$ and similarly for linear and second-order conditional averages, that is $\overline{h}_{\mathrm{T},1}^L=\overline{h}_{\mathrm{T}}^L$ and $\overline{h}_{\mathrm{T},1}^{(\mu)}=\overline{h}_{\mathrm{T}}^{(\mu)}$.

Drawing on the Boccotti (2000) distribution for wave heights, the third-order nonlinear threshold exceeded with probability $q$ by the maximum wave height at a point is given by
\begin{equation} H_{\mathrm{T},q}=h_{\mathrm{T},q}^{(\Lambda)}\sqrt{2(1+\psi^{*})},
\end{equation}
and the conditional mean
\begin{equation} \overline{H}_{\mathrm{T},q}=\overline{h}_{\mathrm{T},q}^{(\Lambda)}\sqrt{2(1+\psi^{*})}.
\end{equation}
Here, the threshold $h_{\mathrm{T},q}^{(\Lambda)}$ accounts for kurtosis effects only and it follows from Eq. (\ref{PNL1}) with $\mu=0$. Similarly, the conditional mean $\overline{h}_{\mathrm{T},q}^{(\Lambda)}$ follows from Eq. (\ref{h1n}). %As $q$ tends to zero $H_{q}$ tend to coincide with the mean $\overline{H}_{q}$.  
For Gaussian waves, the linear counterparts are given by
\begin{equation} H_{\mathrm{T},q}^{(L)}=h_{\mathrm{T},q}^{(L)}\sqrt{2(1+\psi^{*})}
\end{equation}
and
\begin{equation} \overline{H}_{\mathrm{T},q}^{(L)}=\overline{h}_{\mathrm{T},q}^{(L)}\sqrt{2(1+\psi^{*})}.
\end{equation}
Note that $\overline{H}_{\mathrm{T},q=1}$ coincides with the mean maximum nonlinear wave height  $\overline{H}_{\mathrm{T}}$ and the linear conditional mean $\overline{H}_{\mathrm{T},1}^L=\overline{H}_{\mathrm{T}}^L$.

The statistical interpretation of the threshold $h_{q}$ and conditional mean $\overline{h}_{q}$ is as follows. Consider $M$ realizations of a stationary and homogeneous sea state of duration $D$. On this basis, there would be $M$ samples, say $(\eta_{\max}^{(1)},...,\eta_{\max}^{(M)})$ of the maximum surface height $\eta_{\max}$ observed within an area $A$ during the time interval $D$. Then, the threshold $h_{q}$ is exceeded in only $q\thinspace M$ realizations of the ensemble of $M$ sea states, and $\overline{h}_{q}$ is the average of the $q\thinspace M$ largest values in the sample. Similar interpretation holds for the threshold $H_{q}$ and conditional mean $\overline{H}_{q}$ for the maximum wave height.

For the time statistics at a point, $h_{\mathrm{T},q}$ coincides with the standard threshold $c_{n}$ exceeded by the crest heights of the $1/n$ fraction of largest waves if we set $n=N_{\mathrm{D}}/q$, where $N_{\mathrm{D}}$ is the average number of zero-crossing waves expected in a $D$-hour sea state. Similarly, $\overline{h}_{\mathrm{T},q}$ is the crest height average $c_{1/n}$ of  the $q/N_{\mathrm{D}}$ fraction of largest time waves (see, for example, \cite{TayfunFedele2007}). 
Clearly, the previous interpretation does not hold for  space-time or multidimensional random fields. Indeed, the notion of zero-upcrossings is generalized to that of Euler Characteristics of excursion sets and wave counting is not as obvious as in one-dimensional random processes as there is no clear definition of a space-time wave \citep{Fedele2012}.

\begin{figure}[t]
\centering\includegraphics[width=1.04\columnwidth]{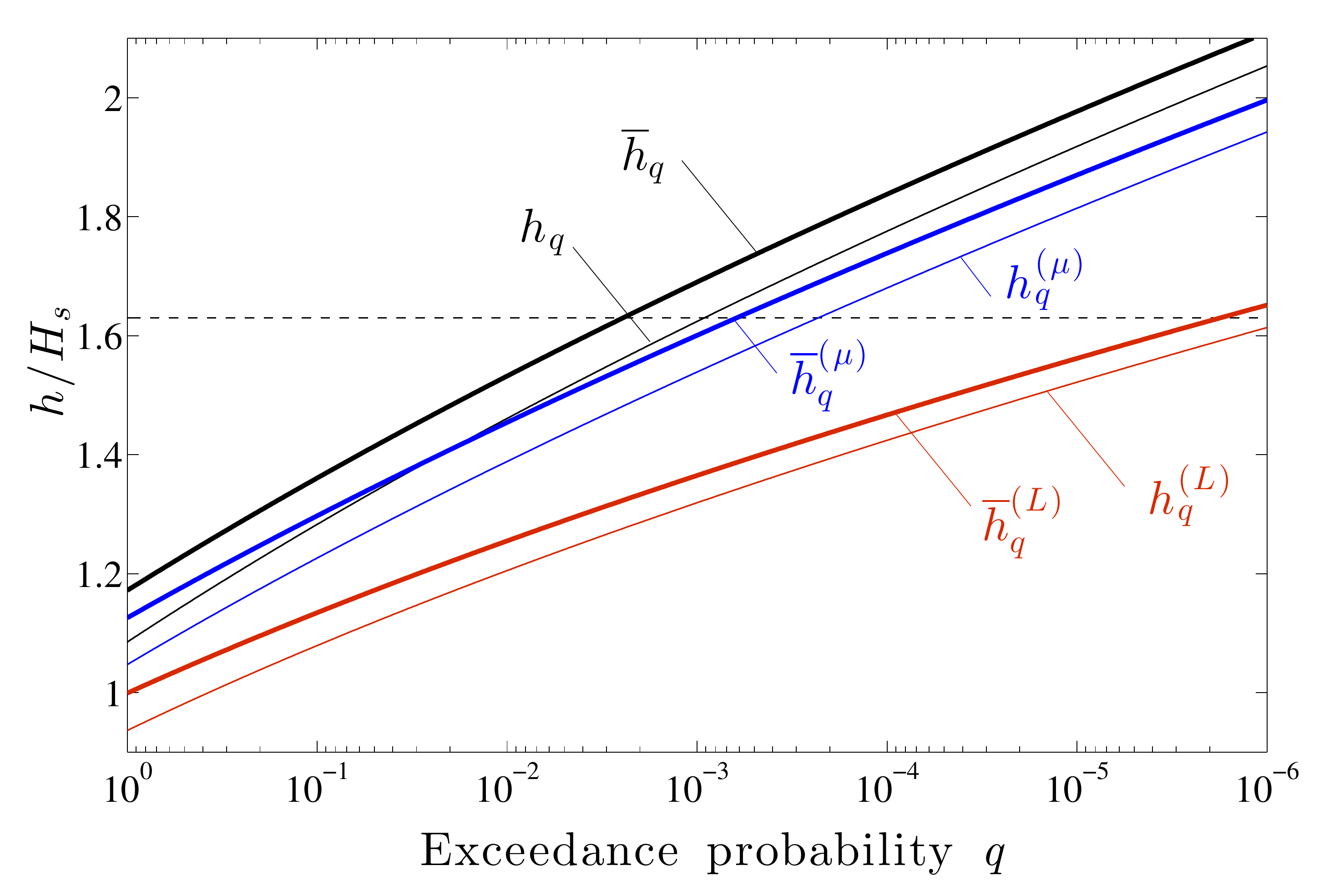}
\caption{Andrea wave: third-order nonlinear threshold $h_{\mathrm{T},q}$ for the maximum crest height at a point as a function of the exceedance probability $q$ for a $3$-hour sea state (thin black line). The linear threshold $h_{\mathrm{T},q}^{(L)}$ (thin red line) and the second-order threshold $h_{\mathrm{T},q}^{(\mu)}$ (thin blue line) are also shown.
The conditional averages $\overline{h}_{q}$, $h_{\mathrm{T},q}^{(\mu)}$ and $\overline{h}_{q}^{(L)}$ are also displayed (bold lines). The horizontal dashed line indicates the actual Andrea wave crest value. Note that $\overline{h}_{\mathrm{T},q=1}$ coincides with the mean maximum surface height  $\overline{h}_{\mathrm{T}}$ and similarly for linear and second-order conditional averages, i.e $\overline{h}_{\mathrm{T},1}^L=\overline{h}_{\mathrm{T}}^L$ and $\overline{h}_{\mathrm{T},1}^{(\mu)}=\overline{h}_{\mathrm{T}}^{(\mu)}$.The wave steepness $\mu\sim0.05$ and the excess kurtosis $\lambda_{40}\sim0.1$ \citep{Bitner_Andrea2014,Dias2015}. }
\label{FIG8} 
\end{figure}

Consider now the hindcasted ERA-interim Andrea sea state during which the rogue wave occurred. Fig.~\ref{FIG8} displays the third-order threshold $h_{\mathrm{T},q}$ from Eq.~\eqref{h1n} for the maximum crest height at a point as a function of the exceedance probability $q$ for a $3$-hour sea state (thin black line). The linear $h_{\mathrm{T},q}^{(L)}$ (thin red line) and second-order threshold $h_{\mathrm{T},q}^{(\mu)}$ (thin blue line) are also shown in the same figure. 
The corresponding conditional averages are also displayed (bold lines). 
If we account for both second and third-order nonlinearities, the actual crest height point measurement observed ($h_{obs}\sim1.63H_{s}$) nearly coincides with $h_{\mathrm{T},1/1000}\sim1.62H_s$. Thus, only in 1 sea state out of the ensemble the maximum surface height exceeds $1.62H_s$. For second-order nonlinearities only, $h_{obs}$ is slightly below $h_{\mathrm{T},1/10,000}^{(\mu)}\sim1.68H_s$ (thin blue line).
%Equivalently, since the average number of waves in a $5$-hour sea state is $N_D\sim1500$, then the %actual $h_{obs}$ is also exceeded by  $1/(100 N_D)\sim1/10^5$ fraction of largest wave crests. In %this regard, the Draupner event can be interpreted as a relatively rare occurrence of weakly non-%Gaussian seas.
Further, $h_{obs}$ exceeds the linear $h_{\mathrm{T},1/100,000}^{(L)}\sim1.52H_s$ and it nearly coincides with $h_{\mathrm{T},1/1,000,000}^{(L)}$ making the Andrea wave event an extremely rare occurrence in Gaussian seas.
%ateshe expected values for the linear mean $\overline{h}_{T,1/n}^{L}$ and associated threshold %$h_{T,n}^{L}$ at a point are also displayed in the left panel of Fig. (\ref{FIG10F}). Clearly, 
%This represents the average of $1/100$ largest maximum surface or crest heights one expects at a %point in an ensemble of 5-hour sea states with similar statistical structure to the Draupner space-%time field.

\begin{figure}[h]
\centering\includegraphics[width=1.04\columnwidth]{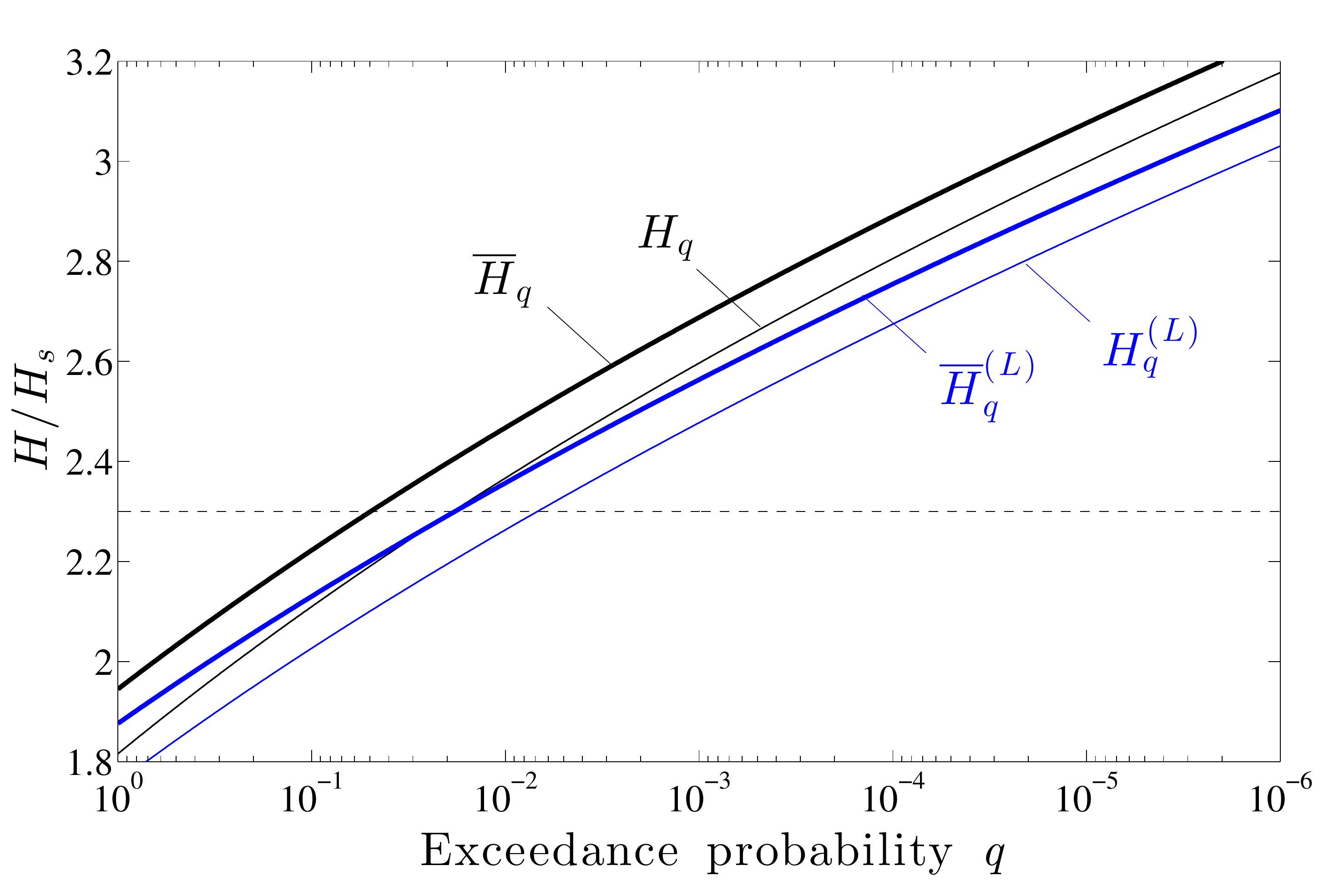}
\caption{Andrea wave: (top) linear and nonlinear thresholds $H_q^{(L)}$ and $H_q$ (thin lines) of the maximum wave height at a point as a function of the exceedance probability $q$ for a $3$-hour sea state. The conditional averages $\overline{H}_{q}^{(L)}$ and $\overline{H}_{q}$ (bold lines) are also shown. The horizontal dashed line indicates the actual Andrea wave height value. Note that $\overline{H}_{\mathrm{T},q=1}$ coincides with the mean maximum wave height  $\overline{H}_{\mathrm{T}}$ and similarly the linear conditional average $\overline{H}_{\mathrm{T},1}^L=\overline{H}_{\mathrm{T}}^L$. The excess kurtosis $\lambda_{40}\sim0.1$ \citep{Bitner_Andrea2014,Dias2015}.}
\label{FIG9} 
\end{figure}

The predictions for the maximum wave height at a point are shown in  Fig. (\ref{FIG9}). Note that the actual wave height observed at Andrea ($H_{obs}\sim2.3H_{s}$) exceeds the third-order threshold $H_{\mathrm{T},1/10}=2.13H_s$ and it nearly coincides with the linear $H_{\mathrm{T},1/100}^{(L)}=2.26H_s$.  This indicates that the Andrea wave most likely occurred in the configuration of a maximum crest and the associated wave height was not the largest and not as extreme as the crest height.  Indeed, according to Boccotti's (2000) QD theory, extreme waves with the largest wave height have almost symmetrical crests and troughs, but the Andrea wave had a strong crest-trough asymmetry (\cite{Magnusson2013}). 

%However, Stokes water wave theory is irrotational and inviscid and the neglected wave breaking is a limiter %on the height of large crests. As a result, higher order Stokes corrections may yield conservative estimates %of the actual extremes. Accounting for wave breaking in the prediction of wave extremes is a challenge and %beyond the scope of this work. 

Note that $h_{T,1/1,000}$ is the same as the threshold $c_{1,000,000}$ exceeded by the crest of one wave in a sample of $n=N_{\mathrm{D}}/q=10^6$ waves, where $q=10^{-3}$ and $N_{\mathrm{D}}\sim10^3$ (mean period $\sim12$ s). Thus, in ideal stationary conditions that last forever the Andrea crest height $h_{obs}$ is exceeded in average every $R=0.5$ years. 
However, oceanic seas are non-stationarity and a long-term analysis of storms is required to estimate the probability of occurrence $P_s$ of a $D$-hour sea state as that similar to the Andrea field. Then, the probability that the threshold $h_q$ is exceeded  during the lifetime span $L$ of an offshore structure is $P=P_s q$, which is an important design parameter.  The associated return period $R$ follows from the relation $P=\mathrm{exp}(-L/R)$, which assumes a Poisson statistics for storm occurrences \citep{Boccotti2000}. Such an analysis requires predicting extremes of non-stationary wave fields following \cite{haver2002} (see also \citep{Boccotti2000, Fedele2012}), but this is beyond the scope of this paper.

In summary, second-order skewness nonlinearities of crest heights are dominant as also indicated by extensive statistical analyses of oceanic wave measurements \citep{TayfunFedele2007,Tayfun2008,Fedele2008a,Fedele2009}. Further, in accord with \cite{Janssen2003}, bound kurtosis effects cannot be neglected and must be accounted for in order to obtain more accurate estimates of both crest and wave height extremes. 

\section{Concluding remarks}

Current freak wave warning systems rely on Janssen's
(2003) theory for the kurtosis of surface elevations, a key result
with significant implications to the understanding of the role of
nonlinear wave interactions in rogue wave formation \citep{Janssen2009}.
The present results suggest that in typical oceanic fields
third-order quasi-resonant interactions do not appear to play a significant
role in the wave growth. Fedele's (2015) refinement of Janssen's theory
 shows that the large excess kurtosis transient observed
during the initial stage of wave evolution is a result of the unrealistic
assumption that the initial wave field is homogeneous and Gaussian.
Oceanic wave fields are typically non homogeneous both in space and
time, and initial conditions become irrelevant as the wave field tends to a more realistic non-Gaussian nonlinear state where statistics are affected mainly by bound harmonics (\cite{Shrira2013_JFM,Shrira2014_JPO}).
%In typical oceanic storms where dominant waves are characterized by
%$\nu\sim0.2-0.4$ and $\sigma_{\theta}\sim0.2-0.4$, this adjustment
%is rapid since the time scale $t_{c}/T_{0}\sim O(1)$ with $T_{0}\sim10-14$
%s and the dynamic kurtosis peak is negligible compared to the bound
%counterpart. 
In this regime, statistical predictions of extreme waves
can be based on the Tayfun (1980) and Janssen (2009) models to account for both second and third-order nonlinearities.
In particular, skewness effects on crest heights are dominant and bound kurtosis must be accounted for to attain more accurate predictions of extremes. %Further studies are desirable to determine the general expressions of fourth order cumulants of the wave surface relying on Janssen's (2009) third-order Hamiltonian formulation, but this is beyond the scope of this paper. 

%Thus, Janssen's theory should be reconsidered to expand its range of validity to 
%in homogeneous seas characterized by broadband spectra.

Our statistical analysis of the Andrea storm reveals that space-time extremes cannot explain the actual large Andrea time crest and wave height observed at the measurement point. The ST analysis simply predicts that the mean maximum sea surface height expected over the Ekofisk platform's area is higher than the mean maximum height expected at a fixed point within the same area, irrespective of the significant wave height level. The actual maximum occurs at some point within the area or its boundaries, and the likelihood of that point's coincidence with the point where measurements are done is essentially nil.

Further, both of the time and space-time averages of maxima underestimate the actual Andrea crest and wave heights observed. If one accounts for both second and third-order nonlinearities, the actual crest height nearly coincides with the threshold $h_{1/1000}$ exceeded with probability $1/1000$ by the maximum crest height at a point in a typical $3$-hour sea state. This suggests that the Andrea rogue wave is likely to be a rare occurrence in quasi-Gaussian seas. 
%dominated by second order nonlinearities.
%Indeed, third order nonlinearities due to bound harmonics are not relevant as the associated kurtosis %$\sim0.1$ in accord with the numerical simulations by \citep{Bitner_Andrea2014,Dias2015}. Indeed, %the second order crest height slightly increases by just 2-3% if one accounts for bound kurtosis effects. 
 Instead, the actual wave height is not so extreme as the crest height as it nearly coincides with the threshold $H_{1/100}$ exceeded with probability $1/100$ by the maximum wave height at a point. The wave height was not the largest as wave measurements indicate a strong crest-trough asymmetry of the Andrea wave, and according to Boccotti's (2000) QD theory this most likely occurred in the configuration of a maximum crest. 
 
Finally, the present conceptual framework for wave extremes can be applied to higher order resolution wave forecasts and operational models for more accurate predictions of rogue waves.

\section{Acknowledgments}

FF is grateful to Jean Bidlot for providing the ERA-interim data of
the Andrea and Draupner storms and for his support in the data analysis. FF also
thanks Michael Banner, Luigi Cavaleri, George Forristall, Aris P. Georgakakos, Peter A. E. M. Janssen, Victor Shrira
and M. Aziz Tayfun for discussions on nonlinear wave statistics and
random wave fields. Further, FF thanks M. Aziz Tayfun and Philip J. Roberts
for revising an early draft of the manuscript, Francesco Barbariol for the support with the analysis of Acqua Alta measurements and Guillermo Gallego for his support with LaTeX. FF acknowledges partial support from NSF grant CMMI-1068624.

%%%%%%%%%%%%%%%%%%%%%%%%%%%%%%%%%%%%%%%%%%%%%%%%%%%%%%%%%%%%%%%%%%%%%
% ACKNOWLEDGMENTS
%%%%%%%%%%%%%%%%%%%%%%%%%%%%%%%%%%%%%%%%%%%%%%%%%%%%%%%%%%%%%%%%%%%%%
%
%\acknowledgments
%Start acknowledgments here.

%%%%%%%%%%%%%%%%%%%%%%%%%%%%%%%%%%%%%%%%%%%%%%%%%%%%%%%%%%%%%%%%%%%%%
% APPENDIXES
%%%%%%%%%%%%%%%%%%%%%%%%%%%%%%%%%%%%%%%%%%%%%%%%%%%%%%%%%%%%%%%%%%%%%
%
% Use \appendix if there is only one appendix.
%\appendix

% Use \appendix[A], \appendix}[B], if you have multiple appendixes.
%\appendix[A]

%% Appendix title is necessary! For appendix title:
%\appendixtitle{}

%%% Appendix section numbering (note, skip \section and begin with \subsection)
% \subsection{First primary heading}

% \subsubsection{First secondary heading}

% \paragraph{First tertiary heading}

\appendix[A]
\appendixtitle{Dynamic and Bound Kurtosis}

For narrowband waves in deep waters, the evolution of the dynamic
excess kurtosis from initial Gaussian conditions is given by (\cite{fedele2014kur})
\begin{equation}
C_{4}^{d}=BFI^{2}J(\tau,R)\label{C4}
\end{equation}
where 
\begin{equation}
J(\tau;R) = 2\mathrm{\, Im}\int_{0}^{\tau}\frac{1}{\sqrt{1-2i\alpha+3\alpha^{2}}\sqrt{1+2iR\alpha+3R^{2}\alpha^{2}}}d\alpha.\label{JR}
\end{equation}
The maximum is attained at $\tau=\tau_{c}$ (see Eq.~\eqref{tau})
and given by
\begin{equation}
C_{4,\max}^{d}(R)=BFI^{2}J_{p}(R),\label{Cmax}
\end{equation}
where
\begin{align*}
J_{p}(R) &= J\left(\frac{1}{\sqrt{3R}};R\right)\\
&=\mathrm{Im}\int_{0}^{\frac{1}{\sqrt{3R}}}\frac{2}{\sqrt{1-2i\alpha+3\alpha^{2}}\sqrt{1+2iR\alpha+3R^{2}\alpha^{2}}}d\alpha,
\end{align*}
and $\mathrm{Im}(a)$ denotes the imaginary part of $a$.

Drawing on \cite{JanssenOnorato2007} and \cite{Janssen2009}, the dynamic kurtosis in intermediate waters of depth $d$ is simply computed by replacing the Benjamin-Feir Index with
\[
BFI_{S}^2=BFI^{2}\alpha_{S}
\]
where the depth factor
\[
\alpha_{S}=-\left(\frac{c_{g}}{c_{0}}\right)^{2}\frac{gX_{nl}}{k_{0}\omega_{0}\omega_{0}^{''}},
\]
depends on the mean frequency $\omega_0$
\[
\omega_{0}=\sqrt{gk_{0}T_{0}},\qquad D_{0}=\mathrm{tanh}(k_{0}d)
\]
corresponding to the mean wavenumber $k_{0}$ via the linear dispersion relation, the group velocity 
\[
c_{g}=\omega_{0}^{'}=\frac{1}{2}c_{0}\left\{ 1+\frac{2k_{0}d}{\mathrm{sinh}(2k_{0}d)}\right\} ,\qquad c_{0}=\frac{\omega_{0}}{k_{0}},
\]
where $\omega_{0}^{'}$ is the first derivative of the angular frequency 
with respect to the wavenumber $k_{0}$, and the second derivative
\[
\omega_{0}^{''}=-g\frac{\left\{ D_{0}-k_{0}d\left(1-D_{0}^{2}\right)\right\} ^{2}+4\left(k_{0}d\right){}^{2}D_{0}^{2}\left(1-D_{0}^{2}\right)}{4\omega_{0}k_{0}D_{0}}.
\]
Further, the nonlinear interaction coefficient 
\[
X_{nl}=\frac{9D_{0}^{4}-10D_{0}^{2}+9}{8D_{0}^{3}}-\frac{1}{k_{0}d}\left\{ 1+\frac{\left(2c_{g}-c_{0}/2\right)^{2}}{c_{s}^{2}-c_{g}^{2}}\right\} ,
\]
where $c_{s}=\sqrt{gd}$ is the phase velocity in shallow waters. 

The NB bound kurtosis $C_{4}^{b}$ is given by \cite{JanssenJFM2009} in the form
\begin{equation}
C_{4}^{b}=\lambda_{40}/3=8\mu_{0}^{2}\left(\beta+\gamma+2(\alpha+\Delta)^{2}\right),\label{Cb}
\end{equation}
 where the wave steepness $\mu_{0}=k_{0}\sigma$, 
\[
\alpha=\frac{3-T_{0}^{2}}{4T_{0}^{3}},\qquad\beta=\frac{3}{64}\frac{8+(1-T_{0}^{2})^{3}}{T_{0}^{6}},\qquad\gamma=-\frac{1}{2}\alpha^{2},
\]
and the wave-induced mean sea level variation 
\[
\Delta=-\frac{1}{4}\frac{c_{S}^{2}}{c_{S}^{2}-c_{g}^{2}}\left[2\frac{1-T_{0}^{2}}{T_{0}}+\frac{1}{x}\right].
\]
In deep waters, $C_{4}^{b}=6\mu_0^{2}$. Note that Eq.~\eqref{Cb} is not valid for small water depths as second and third-order terms of the associated Stokes expansion can be larger than the linear counterpart (see Eq. (A18) in \cite{JanssenJFM2009}) . This implies the constraints $\alpha\mu_m\leq1$ and $\beta\mu_m/\alpha\leq1$, which are not violated for the Andrea sea state. Note that further restrictions apply as occurrences of spurious  crests on the troughs of large, relatively steep second-order Stokes waves are anomalous and not an inherent characteristic of real waves \citep{Tayfun2013}. 

\appendix[B]
\appendixtitle{Space-Time Statistical Parameters}

For space-time extremes, the coefficients in Eq.~\eqref{NST} are given by (\cite{Baxevani2006,Fedele2012})
\[
M_{3}=2\pi\frac{D}{\overline{T}}\frac{\ell_{x}}{\overline{L_{x}}}\frac{\ell_{y}}{\overline{L_{y}}}\alpha_{xyt},
\]
\begin{align*}
M_{2}=&\sqrt{2\pi}\left(\frac{D}{\overline{T}}\frac{\ell_{x}}{\overline{L_{x}}}\sqrt{1-\alpha_{xt}^{2}}\right.\\
&\left.+\frac{D}{\overline{T}}\frac{\ell_{y}}{\overline{L_{y}}}\sqrt{1-\alpha_{yt}^{2}}+\frac{\ell_{x}}{\overline{L_{x}}}\frac{\ell_{y}}{\overline{L_{y}}}\sqrt{1-\alpha_{xy}^{2}}\right),
\end{align*}
%\[
%M_{2}=\sqrt{2\pi}\left(\frac{D}{\overline{T}}\frac{\ell_{x}}{\overline{L_{x}}}\sqrt{1-\alpha_{xt}^{2}}+\frac{D}{\overline{T}}\frac{\ell_{y}}{\overline{L_{y}}}\sqrt{1-\alpha_{yt}^{2}}+\frac{\ell_{x}}{\overline{L_{x}}}\frac{\ell_{y}}{\overline{L_{y}}}\sqrt{1-\alpha_{xy}^{2}}\right),
%\]
\[
M_{1}=N_{D}+N_{x}+N_{y},
\]
where 
\[
N_{D}=\frac{D}{\overline{T}},\qquad N_{x}=\frac{\ell_{x}}{\overline{L_{x}}},\qquad N_{y}=\frac{\ell_{y}}{\overline{L_{y}}}
\]
are the average number of waves occurring during the time interval
$D$ and along the $x$ and $y$ sides of length $\ell_{x}$ and $\ell_{y}$,
respectively. They all depend on the mean period $\overline{T}$,
mean wavelengths $\overline{L_{x}}$ and $\overline{L_{Y}}$ in $x$
and $y$ directions:
\[
\overline{T}=2\pi\sqrt{\frac{m_{000}}{m_{002}}},\quad\overline{L_{x}}=2\pi\sqrt{\frac{m_{000}}{m_{200}}},\quad\overline{L_{y}}=2\pi\sqrt{\frac{m_{000}}{m_{020}}}
\]
and
\[
\alpha_{xyt}=\sqrt{1-\alpha_{xt}^{2}-\alpha_{yt}^{2}-\alpha_{xy}^{2}+2\alpha_{xt}\alpha_{yt}\alpha_{xy}}.
\]
Here, 
\[
m_{ijk}=\iint k_{x}^{i}k_{y}^{j}f^{k}S(f,\theta)dfd\theta
\]
are the moments of the directional spectrum $S(f,\theta)$ and 
\[
\alpha_{xt}\!=\!\frac{m_{101}}{\sqrt{m_{200}m_{002}}}, 
\alpha_{yt}\!=\!\frac{m_{011}}{\sqrt{m_{020}m_{002}}},
\alpha_{xy}\!=\!\frac{m_{110}}{\sqrt{m_{200}m_{020}}}.
\]

\appendix[C]
\appendixtitle{Threshold $h_q$ exceeded with probability $q$ by the maximum surface height}

For 3-D random fields, which are of interest in oceanic applications, the nonlinear probability $P_{\mathrm{FST}}^{(nl)}(\xi;A,D)$ that the maximum surface elevation $\eta_{\max}$ over the area $A$ and during a time interval $D$ exceeds the generic threshold $\xi H_{s}$ is
equal to the probability of exceeding the threshold $\xi_{0}$,
that is 
\begin{equation}
P_{\mathrm{FST}}^{(nl)}(\xi;A,D)=P_{\mathrm{ST}}(\xi_{0};A,D)\left(1+\Lambda\xi_{0}^{2}(4\xi_{0}^{2}-1)\right),\label{Pid1}
\end{equation}
where the Gaussian probability of exceedance (see Eq. \ref{PV} and \citep{adler1981geometry,adler2009random})
\begin{equation}
\begin{split}
P_{\mathrm{ST}}(\xi_{0};A,D)
&=\mathrm{Pr}\left\{ \eta_{\max}>\xi_{0}H_{s}\right\}\\ &=(16M_{3}\xi_{0}^{2}+4M_{2}\xi_{0}+M_{1}\mathrm{\mathrm{)exp}(-8\xi_{0}^{2})},\label{PV1}
\end{split}
\end{equation}
the parameter $\Lambda$ is given in Eq.~\ref{Lambda} and amplitudes $\xi$ and $\xi_0$ are related by the Tayfun (1980) quadratic equation 
\begin{equation}
\xi=\xi_{0}+2\mu\xi_{0}^{2}.\label{sub}
\end{equation}
The nonlinear threshold $h_{q}=\xi_{q} H_s$ exceeded with probability $q$ by $\eta_{\max}$ is given by
\begin{equation}
P_{\mathrm{FST}}^{(nl)}(\xi_q;A,D)=q,\label{PNL}
\end{equation}
or equivalently
\begin{equation}
P_{ST}(\xi_{0,q};A,D)\left(1+\Lambda\xi_{0,q}^{2}(4\xi_{0,q}^{2}-1)\right)=q,\label{EE}
\end{equation}
where from Eq.~\eqref{sub}
\begin{equation}
\xi_{q}=\xi_{0,q}+2\mu\xi_{0,q}^{2}.\label{sub3}
\end{equation}
The mean $\overline{h}_q$ of $\eta_{\max}$ conditioned on $\eta_{\max}\ge h_q$ then follows by definition from
\begin{equation}
\overline{h}_{q}/H_s=\frac{\int_{\xi_{q}}^{\infty}\xi p_{\mathrm{ST}}^{(nl)}(\xi;A,D)d\xi}{\int_{\xi_{q}}^{\infty}p_{\mathrm{ST}}^{(nl)}(\xi;A,D)d\xi},\label{h1nNL}
\end{equation}
where the nonlinear pdf
\begin{equation}
p_{\mathrm{FST}}^{(nl)}(\xi;A,D)=-\frac{dP_{\mathrm{ST}}^{(nl)}}{d\xi}.\label{pdf}
\end{equation}
Integrating~\eqref{h1nNL} by parts yields
\[
\overline{h}_{q}/H_s=\frac{\xi_{q}P_{\mathrm{ST}}^{(nl)}(\xi_{q})+\int_{\xi_{q}}^{\infty}P_{\mathrm{ST}}^{(nl)}(\xi)d\xi}{P_{\mathrm{ST}}^{(nl)}(\xi_q)},
\]
and it follows from Eq.~\eqref{PNL} that
\[
\overline{h}_{q}/H_s=\xi_{q}+\frac{1}{q}\int_{\xi_{q}}^{\infty}P_{\mathrm{ST}}^{(nl)}(\xi)d\xi.
\]
Next, note that 
\begin{equation}
\int_{\xi_{q}}^{\infty}P_{FST}^{(nl)}(\xi)d\xi=\int_{\xi_{q}}^{\infty}P_{ST}\left(\xi_{0}\right)\left(1+\Lambda\xi_{0}^{2}(4\xi_{0}^{2}-1)\right)d\xi,\label{int}
\end{equation}
where $\xi_{0}\left(\xi\right)$ is a function of the nonlinear $\xi$ given by Eq. (\ref{sub}).
A change of variables via $\xi_{0}\left(\xi\right)=y$ and $d\xi=\frac{d\xi}{d\xi_{0}}dy=(1+4\mu y)dy$ simplify the preceding expression to 
\begin{equation}
\int_{\xi_q}^{\infty}P_{\mathrm{ST}}^{(nl)}(\xi)d\xi=\int_{\xi_{0,q}}^{\infty}P_{\mathrm{ST}}(y)F(y)dy,\label{int-1}
\end{equation}
where $F(y)=(1+\Lambda y^2(4y^2-1))\left(1+4\mu y\right)$. Further, $\xi_{0,q}$ follows from Eq.~\eqref{EE} and it relates to $\xi_q$ via Eq.~\eqref{sub3}. 
Therefore,
\[
\overline{h}_{q}/H_s=\xi_{q}+\frac{1}{q}\int_{\xi_{0,q}}^{\infty}P_{\mathrm{ST}}(y)F(y)dy.
\]
This expression is written in the form
\begin{equation}
\overline{h}_{q}=h_{q}+\left(\chi_{1,q}+\mu\chi_{2,q}+\Lambda\chi_{3,q}+\mu\Lambda\chi_{4,q}\right)H_s/q,\label{hqmean}
\end{equation}
where
\begin{align*}
\chi_{1,q}&=\int_{\xi_{0,q}}^{\infty}P_{ST}(\xi)d\xi\\
&=p_{1,q}\exp(-8\xi_{0,q}^{2})+\frac{\sqrt{2\pi}}{8}(M_{1}+M_{3})\mathrm{Erfc}(2\sqrt{2}\xi_{0,q}),\\
\chi_{2,q}&=\int_{\xi_{0,q}}^{\infty}4yP_{ST}(y)dy\\
&=p_{2,q}\exp(-8\xi_{0,q}^{2})+\frac{\sqrt{2\pi}}{8}M_{2}\mathrm{Erfc}(2\sqrt{2}\xi_{0,q}),\\
\chi_{3,q}&=\int_{\xi_{0,q}}^{\infty}y^{2}\left(4y^{2}-1\right)P_{ST}(y)dy\\
&=\mathrm{\mathit{p}_{3,\mathit{q}}exp}(-8\xi_{0,q}^{2})-\frac{M_{1}-3M_{3}}{512}\sqrt{2\pi}\mathrm{Erfc}(2\sqrt{2}\xi_{0,q}),\\
\chi_{4,q}&=\int_{\xi_{0,q}}^{\infty}4y^{3}\left(4y^{2}-1\right)P_{ST}(y)dy\\
&=p_{4,q}\exp(-8\xi_{0,q}^{2})+\frac{3\sqrt{2\pi}}{512}M_{2}\mathrm{Erfc}(2\sqrt{2}\xi_{0,q}),
\end{align*}
$\mathrm{Erfc}(x)$ is the complementary error function and 
\begin{align*}
p_{1,q}=&\frac{1}{4}\left(M_{2}+4M_{3}\xi_{0,q}\right),\\
p_{2,q}=&\frac{1}{4}\left(M_{1}+2M_{3}+4M_{2}\xi_{0,q}+16M_{3}\xi_{0,q}^{2}\right),\\
p_{3,q}=&\xi_{0,q}\left(\frac{3M_{3}-M_{1}}{64}+\frac{M_{1}+M_{3}}{4}\xi_{0,q}^{2}+M_{2}\xi_{0,q}^{3}+4M_{3}\xi_{0,q}^{4}\right),\\
p_{4,q}=&\frac{M_{3}}{16}+\frac{3M_{2}}{64}\xi_{0,q}+\frac{M_{3}}{2}\xi_{0,q}^{2}+\frac{M_{2}}{4}\xi_{0,q}^{3}+\\
&(M_{1}+2M_{3})\xi_{0,q}^{4}+4M_{2}\xi_{0,q}^{5}+16M_{3}\xi_{0,q}^{6}.
\end{align*}
%% Important!
%\appendcaption{<appendix letter and number>}{<caption>} 
%must be used for figures and tables in appendixes, e.g.,
%
%\begin{figure}
%\noindent\includegraphics[width=19pc,angle=0]{figure01.pdf}\\
%\appendcaption{A1}{Caption here.}
%\end{figure}
%
% All appendix figures/tables should be placed in order AFTER the main figures/tables, i.e., tables, appendix tables, figures, appendix figures.
%
%%%%%%%%%%%%%%%%%%%%%%%%%%%%%%%%%%%%%%%%%%%%%%%%%%%%%%%%%%%%%%%%%%%%%
% REFERENCES
%%%%%%%%%%%%%%%%%%%%%%%%%%%%%%%%%%%%%%%%%%%%%%%%%%%%%%%%%%%%%%%%%%%%%
% Make your BibTeX bibliography by using these commands:
\bibliographystyle{ametsoc2014}
\bibliography{biblioFranco}

%%%%%%%%%%%%%%%%%%%%%%%%%%%%%%%%%%%%%%%%%%%%%%%%%%%%%%%%%%%%%%%%%%%%%
% TABLES
%%%%%%%%%%%%%%%%%%%%%%%%%%%%%%%%%%%%%%%%%%%%%%%%%%%%%%%%%%%%%%%%%%%%%
%% Enter tables at the end of the document, before figures.
%%
%
%\begin{table}[t]
%\caption{This is a sample table caption and table layout.  Enter as many tables as
%  necessary at the end of your manuscript. Table from Lorenz (1963).}\label{t1}
%\begin{center}
%\begin{tabular}{ccccrrcrc}
%\hline\hline
%$N$ & $X$ & $Y$ & $Z$\\
%\hline
% 0000 & 0000 & 0010 & 0000 \\
% 0005 & 0004 & 0012 & 0000 \\
% 0010 & 0009 & 0020 & 0000 \\
% 0015 & 0016 & 0036 & 0002 \\
% 0020 & 0030 & 0066 & 0007 \\
% 0025 & 0054 & 0115 & 0024 \\
%\hline
%\end{tabular}
%\end{center}
%\end{table}

%%%%%%%%%%%%%%%%%%%%%%%%%%%%%%%%%%%%%%%%%%%%%%%%%%%%%%%%%%%%%%%%%%%%%
% FIGURES
%%%%%%%%%%%%%%%%%%%%%%%%%%%%%%%%%%%%%%%%%%%%%%%%%%%%%%%%%%%%%%%%%%%%%
%% Enter figures at the end of the document, after tables.
%%
%
%\begin{figure}[t]
%  \noindent\includegraphics[width=19pc,angle=0]{figure01.pdf}\\
%  \caption{Enter the caption for your figure here.  Repeat as
%  necessary for each of your figures. Figure from \protect\cite{Knutti2008}.}\label{f1}
%\end{figure}

\end{document}